\documentclass[superscriptaddress,prd,aps,preprint,eqsecnum,floats]{revtex4}
\usepackage{epsfig}
\usepackage{amsfonts}
\usepackage{amssymb}
\usepackage{amsmath}
\usepackage{bm}
\usepackage{bbold}

  \newcommand \Sc {\stackrel{\circ}{\mathcal{S}}}

  \newcommand\beq{\begin{equation}}
  \newcommand\eeq{\end{equation}}
  \newcommand\beqn{\begin{eqnarray}}
  \newcommand\eeqn{\end{eqnarray}}
  \newcommand{\la}{\langle}
  \newcommand{\ra}{\rangle}

\newcommand{\gb}[1]{\mbox{\boldmath ${#1}$ }}
\def\pnul{\raise-.3ex\hbox{$\stackrel{\circ}{p}$}}\relax
\def\snul{\raise-.3ex\hbox{$\stackrel{\circ}{s}$}}\relax
\def\half{\mbox{\small $\frac{1}{2}$}}\relax

\begin{document}
\title{\begin{flushright}{\small BNL-HET-04/6\\ \vspace{-.15in} \small IC/HEP/04-2\vspace{.25in}\\   \endflushright}\Large  \bf{A critique of the angular momentum sum rules and \vspace{-.15in}\begin{center} a new angular momentum sum rule \end{center}}}
\normalsize
\author{B.L.G. Bakker}
\affiliation{Department of Physics and Astronomy, Vrije
Universiteit, Amsterdam}
\author{E. Leader}
\affiliation{Imperial College London, SW7 2BW}
\author{T. L. Trueman}
\affiliation{Physics Department, Brookhaven National Laboratory,
Upton, NY 11973}
\date{\small June 10, 2004}

\vspace{3cm} Submitted to Physical Review D

\begin{abstract}
We present a study of the tensorial structure of the hadronic
matrix elements of the angular momentum operators $\bm{J}$. Well
known results in the literature are shown to be incorrect, and we
have taken pains to derive the correct expressions in three
different ways, two involving explicit physical wave packets and
the third, totally independent, based upon the rotational
properties of the state vectors. Surprisingly it turns out that
the results are very sensitive to the type of relativistic spin
state used to describe the motion of the particle i.e.~whether a
canonical (i.e.~boost) state or a helicity state is utilized. We
present results for the matrix elements of the angular momentum
operators, valid in an arbitrary Lorentz frame, both for helicity
states and canonical states.

These results are relevant for the construction of angular
momentum sum rules, relating the angular momentum of a nucleon to
the spin and orbital angular momentum of its constituents. It
turns out that it is necessary to distinguish carefully whether
the motion of the partons is characterized via canonical or
helicity spin states. Fortunately, for the simple parton model
interpretation, when the proton moves along $OZ$, our results for
the sum rule based upon the matrix elements of $J_z$ agree with
the often used sum rule found in the literature. But for the
components $J_x, J_y$ the results are different and lead to a new
and very intuitive sum rule for transverse polarization.
\end{abstract}
\maketitle

\newpage
\section{Introduction and summary of results}
\label{sec.1}
Sum rules, relating the total angular momentum of a nucleon to the
spin and orbital angular momentum carried by its constituents, are
interesting and important in understanding the internal structure
of the nucleon. Indeed it is arguable that the main stimulus for
the tremendous present day experimental activity in the field of
spin-dependent structure functions was the surprising result of
the European Muon Collaborations polarized DIS experiment in 1988
\cite{ref.01}, which, via such sum rules, led to what was called a
``spin crisis in the parton model'' \cite{ref.02}, namely the
discovery that the spins of its quarks provide a very small
 contribution to the angular momentum of the proton. A
key element in deriving such sum rules is a precise knowledge of
the tensorial structure of the expectation values of the angular
momentum operators $J_i$ in a state $| p\, ,\sigma\ra$ of the
nucleon, labeled by its momentum $p$, and with some kind of
specification of its spin state,  denoted here non-commitally by
$\sigma$.

In a much cited paper \cite{ref.03}, Jaffe and Manohar stressed
the subtleties involved in deriving general angular momentum sum
rules. As they point out, too naive an approach leads immediately
to highly ambiguous divergent integrals, and a careful limiting
procedure has to be introduced in order to obtain physically
meaningful results.  In this it  is essential to work with
non-diagonal matrix elements $\la p', \sigma|\gb{J} | p\,
,\sigma\ra$ and, as discussed below, this can have some unexpected
consequences. Jaffe and Manohar comment that to justify rigorously
the steps in such a procedure requires the use of normalizable
wave packets, though they do \textit{not} do this explicitly in
their paper.

In a later paper \cite{ref.04}, Shore and White utilized the
approach of Ref.~\cite{ref.03}, including an explicit treatment
with wave-packets, to derive some far reaching conclusions about
the r\^{o}le  of the axial anomaly in these sum rules.

We shall argue that despite all the care and attention to
subtleties, there are flaws in the analysis in \cite{ref.03} and
the results presented there are not entirely general.  Indeed
there are cases where the results of \cite{ref.03} are incorrect.
This, in turn, throws doubt upon some of the conclusions reached
in \cite{ref.04}, which we will examine.

The bulk of our analysis is based on a straightforward wave-packet
approach. However, as we explain, this is rather subtle for
particles with non-zero spin. The key points underlying our
results are:

   (1) Our wave packets are strictly physical, i.e. a superposition
of physical plane-wave states.  This requirement turns out to be
incompatible with some of the steps in \cite{ref.03}.\\

   (2) We give a careful treatment of the Lorentz covariance properties of the
matrix elements involved in the subsidiary steps of the anaylsis.
This leads to tensorial structures which differ in some cases from
those in \cite{ref.03}.\\

  (3) Because our results differ from \cite{ref.03} we have looked
for and found a totally independent way to check our results. This
does not use wave packets and is based upon the transformation
properties of momentum states under rotations. This very direct
approach holds for arbitrary spin, whereas in the wave packet
treatment we are only able to deal with spin $\frac{1}{2}$
particles. It also brings to light some peculiar and unintuitive
properties of helicity states, and this must be taken into account
when deriving spin sum rules. This is important since
we have to deal with gluons in our sum rules. \\

Our results for the matrix elements of $ \gb{J}$ are as follows:

For a massive particle of spin $ \frac{1}{2}$  with 4-momentum $p$
in a canonical spin state $|p,\bm{s}\ra$ (i.e. in a `boost' state
of the kind generally used in textbooks on Field Theory e.g. in
Bjorken and Drell \cite{Bjorken}, or in Peskin and Schroeder
\cite{Peskin} ) , where $ \bm{s}/2$ is the spin eigenvector in the
rest frame $(\bm{s}^{2}=1)$, we show that, for the forward matrix
elements,
\begin{equation}
\la p', \gb{s}|J_i|p, \gb{s}\ra = 2 p^0(2\pi)^{3}\left[
\frac{1}{2} \;s_i + i (\gb{p \times \nabla_{p}})_i  \right]
\delta^{3}\gb{(p'-p)}. \label{eq.5.011}
\end{equation}
The states $|p, \bm{s} \rangle $ are normalized conventionally to
\begin{equation}
\la p',\gb{s}| p, \gb{s}\ra = 2 p^0(2 \pi)^{3} \delta^{3} (\gb{p'-
p}) \label{eq.5.021}
\end{equation}
and we note that in the rest frame
\begin{equation} \label{5.022}
 |\, 0,\bm{s}  \ra = \mathcal{D}^{1/2}_{m 1/2} (R(\bm{s})) |\,
{0,m}\ra
\end{equation}
where $|\, 0,m \ra $ has spin projection $m$ along the
$z$-direction in the rest frame and $R(\bm{s})$ rotates a unit
vector in the $z$-direction into $\bm{s}$ by first a rotation
about $y$ and then a rotation about $z$.

For the purpose of deriving sum rules our result for the matrix
elements \textit{non-diagonal }in the spin label is actually more
useful, namely, for a spin $\half$ particle
  \beqn
   \la p',m'|J_i|p,m \ra
   &=&2p_0 (2\pi)^3 \left[ \frac{1}{2} \sigma_i + i\epsilon_{ijk} p_j
\frac{\partial}{\partial p_k}
\right]_{m'm}\delta^{(3)}(\gb{p'-p}). \label{J11}
   \eeqn
The generalization of these results for arbitrary spin is given in
Eq.~(\ref{JK}).

 Helicity states are more suitable for massless particles such as gluons.
 Using the Jacob-Wick conventions for helicity states \cite{JacobWick} we
find a surprisingly different result, namely,
\begin{equation}
 \langle p',\lambda' | J_i | p,\lambda \rangle = (2\pi)^3 2p_0
 \left[ \lambda \eta_i (\bm{p}) + i (\bm{p} \times \bm{\nabla}_{\bm{p}})_i \right]
 \delta^3(\bm{p}' - \bm{p}) \delta_{\lambda\lambda'}
\label{eq.5.041}
\end{equation}
where
 \beq
\eta_{x}= \cos(\phi)\tan(\theta/2),\qquad \eta_{y}= \sin(\phi)
\tan(\theta/2), \qquad \eta_{z}=1. \label{helicity21}
 \eeq
and $(\theta, \phi)$ are the polar angles of $\gb{p}$.

The first term in Eq.~(\ref{eq.5.011}) differs from the results of
Jaffe and Manohar [3].  If we rewrite their expression (see
Eq.~(\ref{eq.37})) in terms of the independent vectors $\bm{p}$
and $\bm{s}$, we find, for the \textit{expectation} value
\begin{equation}
 \la J_i\ra_{JM} = \frac{1}{4Mp^0} \left\{(3p_0^{2}- M^{2}) s_i
 -\frac{3p_0 +M}{p_0+ M} (\bm{p} \cdot \bm{s}) p_i \right\}
\label{eq.5.051}
\end{equation}
to be compared to
\begin{equation}
\la J_i\ra =\half  s_i \label{eq.5.061}
\end{equation}
arising from the first term in Eq.~(\ref{eq.5.011}).  In general
these are different. However, one may easily check that if $\bm{s}
= \hat{\bm{p}}$ the Jaffe-Manohar value agrees with
Eq.(\ref{eq.5.061}), while if $\bm{s}  \perp \hat{\bm{p}}$ they
are not the same.

The agreement for  $\bm{s} = \hat{\bm{p}}$  is consistent with the
much used and intuitive sum rule
\begin{equation}
\half = \half \Delta \Sigma + \Delta G +
 \langle L^q \rangle + \langle L^G \rangle
\label{eq.5.071}
\end{equation}
In the case that $\bm{s}  \perp \hat{\bm{p}}$ we find a new sum
rule. For a proton with transverse spin vector $\bm{s}_T$ we find
\begin{equation}\label{eq.T1}
\half = \half\, \sum_{q, \,\bar q }\, \int dx \, \Delta _T q^a (x)
+ \sum_{q, \, \bar q, \, G }\langle L_{\bm{s}_T} \rangle^a
\end{equation}
where $L_{\bm{s}_T}$ is the component of $\bm{L}$ along
$\bm{s}_T$. The structure functions $\Delta_T q^a (x) \equiv
h^q_1(x)$ are known as the quark transversity or transverse spin
distributions in the  nucleon. Note that no such parton model sum
rule is possible with the Jaffe-Manohar formula because, as $p
\rightarrow \infty$, Eq.~(\ref{eq.5.051}) for $i=x,y$ diverges.

The result Eq.~(\ref{eq.T1}) has a very intuitive appearance, very
similar to Eq.~(\ref{eq.5.071}).

The organization of our paper is as follows: In Section~2 we
explain why the calculation of the angular momentum matrix
elements is so problematical. Because of the unexpected
sensitivity of the matrix elements of $\gb{J}$ to the type of spin
state used, and because we are forced to use helicity states for
gluons, Section~3 presents a resum\'{e} of the difference between
Jacob-Wick helicity states and canonical ( i.e as in
Bjorken-Drell) spin states. Further, given that our results
disagree with one of the classic papers in the literature, we have
felt it incumbent to summarize, in Section~4, the treatments of
Jaffe-Manohar and Shore-White, pointing out the incorrect steps in
these derivations.

In Section~5 we present a detailed wave-packet derivation of the
structure of the matrix elements of $\bm{J}$ for spin
$\frac{1}{2}$, first for a relativistic  Dirac particle, then in a
field theoretic treatment. We comment here on the claims made in
Shore-White on the role of the axial anomaly in the structure of
the matrix elements. Sections~4 and 5 are heavy going, and the
reader only interested in a quick and direct derivation of the key
results should skip these and read Sections~6, 6.1, 6.3 and 7.

In Section~6 we confirm the results of Section~5 in a completely
independent approach, which is valid for \textit{arbitrary} spin,
based on the rotational properties of canonical and helicity spin
states. We also prove that our results are in conformity with the
demands of Lorentz invariance.

 In Section~7 we derive the most general form of an angular momentum
 sum rule for a nucleon and show that it reduces to the standard, intuitive, sum
 rule for $J_{z}$ when the nucleon is moving along $OZ$. We also
 derive a new sum rule for a transversely polarized nucleon.

 \section{\textsc{The origin of the problem}}
\label{origin}
In the standard approach one relates the matrix elements of the
angular momentum operators to those of the energy-momentum tensor.

Let $T^{\mu\nu}(x)$ be the total energy-momentum density which is
conserved
\begin{equation}
   \partial_\mu T^{\mu\nu}(x) =0.
\label{eq.01}
\end{equation}
Later we shall distinguish between the conserved {\em canonical}
energy-momentum tensor $T^{\mu\nu}_{\rm C}$, which emerges from
Noether's theorem, and which is, generally, not symmetric under
$\mu \leftrightarrow \nu$, and the {\em symmetrised} Belinfante
tensor $T^{\mu\nu}$, which for QCD is given by
\begin{equation}
    T^{\mu\nu}(x) = \half
   \left( T^{\mu\nu}_{\rm C}(x) + T^{\nu\mu}_{\rm C}(x) \right)
\label{eq.02}
\end{equation}
and which is also conserved. For the moment, however, this
distinction is irrelevant.

Being a {\em local} operator, $T^{\mu\nu}(x)$ transforms under
translations as follows
\begin{equation}
    T^{\mu\nu}(x) = e^{i {x \cdot P}} T^{\mu\nu}(0)  e^{-i
{x \cdot P}} , \label{eq.03}
\end{equation}
where the $P^{\mu}$ are the total momentum operators of the
theory.

By contrast the various angular momentum density operators which
are of interest, the orbital angular momentum densities
\begin{equation}
   M^{\mu\nu\lambda}_{\rm orb} (x) \equiv
   x^\nu T^{\mu\lambda}_{\rm C}(x) - x^\lambda T^{\mu\nu}_{\rm C}(x)
\label{eq.04}
\end{equation}
or the version constructed using the symmetrised stress-energy
tensor,
\begin{equation}
   M^{\mu\nu\lambda} (x) \equiv
   x^\nu T^{\mu\lambda}(x) - x^\lambda T^{\mu\nu}(x)
\label{eq.05}
\end{equation}
are not local operators (we shall call them {\em compound})  and
do not transform  according to Eq.~(\ref{eq.03}).

Note that, strictly speaking, the operators relevant to the
angular momentum are the components $M^{0ij}$ where $i,j$ are
spatial indices. However, for reasons of simplicity in utilising
the Lorentz invariance of theory, the authors of
Ref.~\cite{ref.03} prefer to deal covariantly with the entire
tensor $M^{\mu\nu\lambda}$. We shall loosely refer to them also as
angular momentum densities.

The total angular momentum density is
\begin{equation}
   J^{\mu\nu\lambda}(x) =
   M^{\mu\nu\lambda}_{\rm orb} (x) + M^{\mu\nu\lambda}_{\rm spin} (x),
\label{eq.06}
\end{equation}
where the structure of $M^{\mu\nu\lambda}_{\rm spin}$ depends on
the type of  fields involved. From Noether's theorem
$J^{\mu\nu\lambda}(x)$ is a set of conserved densities, i.e.,
\begin{equation}
   \partial_{\mu} J^{\mu\nu\lambda}(x) = 0.
\label{eq.07}
\end{equation}

As a consequence of the densities being conserved, it follows that
the total momentum operators
\begin{equation}
P^\nu \equiv \int d^3 x T^{0\nu} (x) \label{eq.08}
\end{equation}
and the total angular momentum operators $\bm{J}$,
\begin{equation}
   J_z = J^3 = J^{12}, \quad ({\rm cyclical})
\label{eq.09}
\end{equation}
with
\begin{equation}
   J^{ij} \equiv \int d^3 x J^{0ij} (x)
\label{eq.10}
\end{equation}
are conserved quantities, independent of time.

The relationship between the $M^{\mu\nu\lambda} (x)$, constructed
using the symmetrical energy-momentum density and the
$J^{\mu\nu\lambda}(x)$ constructed from the canonical
energy-momentum tensor is extremely interesting and will be
commented on later in Section~4.2. One can show (see e.g.
Ref.~\cite{ref.04}) that
\begin{equation}
   M^{0ij} (x) = J^{0ij}(x) +
        [{\rm E~of~M~terms}] + [{\rm divergence~terms}].
\label{eq.11}
\end{equation}
The [E of M terms] vanish if it is permissible to use the
equations of motion of the theory. The [divergence terms] are of
the form $\partial_\alpha F^{\alpha 0ij} (x)$.

As mentioned we shall be primarily interested in the expectation
values of the physical operators, i.e. in their forward matrix
elements. If $F^{\alpha 0ij} (x)$ were a \underline{local}
operator, it would follow directly that the forward,
momentum-space, matrix elements of the divergence terms in
Eq.~(2.11) vanish. But it is not a local operator. Nonetheless, a
careful treatment using wave packets \cite{ref.04} demonstrates
that the forward matrix elements do indeed vanish. See Section~4
below.

Dropping, as is customary, the [E of M terms], we shall thus
assume the validity of
\begin{equation}
   \langle p,\sigma | \int d^3 x M^{0ij} (\bm{x},0) | p,\sigma \rangle =
   \langle p,\sigma | \int d^3 x J^{0ij}(\bm{x},0) | p,  \sigma \rangle.
\label{eq.2.12}
\end{equation}
We shall return to this question in Sections~5.1 and 5.2.

Of primary interest are the matrix elements of the angular
momentum operators $J^k$ or, equivalently, the $J^{ij}$. Consider
the forward matrix element, at $t=0$,
\begin{eqnarray}
   {\cal M}^{0ij} (p,\bm{s}) & \equiv &
   \langle p,\bm{s} | \int d^3 x M^{0ij}(\bm{x},0) | p,\bm{s} \rangle \\
   & = & \int d^3 x \langle p, \bm{s} | x^i T^{0j} (x) - x^j T^{0i} (x)
| p, \bm{s} \rangle
   \nonumber \\
   & = &  \int d^3 x\,
   x^i \,\langle p, \bm{s} | e^{i {P \cdot x}} T^{0j} (0) e^{-i
{P \cdot x}} | p, \bm{s} \rangle
   - (i\leftrightarrow j)
   \nonumber \\
   & = &  \int d^3 x \, x^i  \,\langle p, \bm{s} | T^{0j} | p, \bm{s} \rangle
    - (i\leftrightarrow j).
\label{ambiguous}
\end{eqnarray}
The integral in Eq.~(\ref{ambiguous}) is totally ambiguous, being
either infinite or, by symmetry, zero.

The essential problem is to obtain a sensible physical expression,
in terms of $p$ and $\bm{s}$, for the above matrix element. The
fundamental idea is to work with a non-forward matrix element and
then to try to approach the forward limit. This is similar to what
is usually done when dealing with non-normalizable plane wave
states and it requires the use of wave packets for a rigorous
justification.

It will turn out that the results are sensitive to the type of
relativistic spin state employed, so in Section~3 we present a
brief resum\'{e} of the properties of relativistic spin states. We
then proceed to discuss the approaches of \cite{ref.03} and
\cite{ref.04} in Section~4, where we shall comment on the dubious
steps in these treatments. The most crucial error in these
treatments is the mishandling of the matrix elements of a
covariant tensor operator. If $T^{\mu\lambda}$ transforms as a
second-rank tensor its \emph{non-forward} matrix elements do
\emph{not} transform covariantly. This was the motivation, decades
ago, for Stapp to introduce $M$-functions~\cite{stapp}. Namely,
the covariance is spoilt, for canonical spin states by the Wigner
rotation, and, for helicity states by the analogous Wick helicity
rotation~\cite{Wick}. Only by first factoring out the
wave-functions (in our case Dirac spinors) i.e. by writing
 \beq
 \la{ p', \mathcal{S}'}|T^{\mu\lambda}|{ p,  \mathcal{S}}\ra  =
\bar{u}(p', \mathcal{S}') \mathcal{T}^{\mu \nu}(p',p) u(p,
\mathcal{S}) .\label{tensorform1}
 \eeq
does the remaining $M$-function, in this case $\mathcal{T}^{\mu
\nu}(p',p)$,  transform covariantly. For local operators the
transformations of the spinors $u$ and $\bar{u}$ cancel between
themselves for forward matrix elements and so the result does have
the naively expected tensor expansion. This is not true in general
for compound operators, in particular the angular momentum and
boost operators.

\section{RELATIVISTIC SPIN STATES}
The definition of a spin state for a particle in motion, in a
relativistic theory, is non-trivial, and is convention dependent.
Namely, starting with the states of a particle at rest, which we
shall denote by $| 0, m\ra$, where  $m$ is the spin projection in
the $z$-direction, one defines states $|p, \sigma\ra$ for a
particle with four-momentum $p$ by acting on the rest frame states
with various boosts and rotations, and the choice of these is
convention-dependent.  The states are on-shell so $p^2=M^2$

There are three conventions in general use  \cite{Leader}

a)  Canonical or boost states as used e.g. in Bjorken and Drell
\cite{Bjorken} or Peskin and Schroeder \cite{Peskin}
\begin{equation}
| p, m \ra = B (\bm{v}) | 0, m \ra \label{pmbo}
\end{equation}
where $B(\bm{v})$ is a pure boost along $\bm{v} = \bm{p}/p_0$, and
$\gb{p}= (p,\theta ,\phi )$ denotes the three-vector part of
$p^\mu$

b)  Jacob-Wick helicity states \cite{JacobWick}
\begin{equation}
|p, \lambda \ra_{JW} = R_{z} (\phi)\, R_y (\theta)\, R_z(-\phi)\,
B_{z} (v) | 0, m = \lambda \ra \label{lbdj}
\end{equation}
where $B_{z}$ is a boost along $OZ$, and the later introduced,
somewhat simpler

c)  Wick helicity states \cite{Wick}
\begin{equation}
| p, \lambda \ra = R_{z}(\phi)\, R_{y} (\theta) \,B_{z}(v) | 0, m
= \lambda \ra \label{przo}
\end{equation}

 From the canonical states of spin-$ \frac{1}{2}$ one can construct the states
\begin{equation}
| p, \bm{s} \ra = B (\bm{v}) | 0, \bm{s} \ra = B
(\bm{v})\mathcal{D}^{\, 1/2}_{m 1/2} (R(\bm{s})) |{0,m}\ra
\label{lpos}
\end{equation}
which, in the rest frame, are eigenstates of spin with spin
eigenvector along the unit vector $\bm{s}$. (The rotation
$R(\bm{s})$ was explained after Eq.~(\ref{5.022})).

The canonical states, with their reference to a rest frame, are
clearly not suitable for massless particles like gluons. Helicity
states, on the other hand, can be used for both massive and
massless particles. However, it turns out that the results for the
canonical states are much more intuitive, so we will generally use
them for $M\neq 0$.

The reason we are emphasizing this distinction between canonical
and helicity states is that the matrix elements of the angular
momentum operators between helicity states are quite bizarre!
Since, for arbitrary $p$, helicity states are just linear
superpositions of canonical states, one may wonder why this is so.
It results from the facts i) that the coefficients in the linear
superposition are $p$-dependent, i.e. depend upon the polar angles
of $\bm{p}$ and ii) that the matrix elements of the angular
momentum operators contain derivatives of $\delta$-functions, and
these, as usual, must be interpreted in the sense of partial
integration, i.e.
\begin{equation}
f(p,p') \frac{\partial}{\partial p_i} \delta^3(p-p') = -
\delta^3(p-p')  \frac{\partial}{\partial p_i}f(p,p')
\end{equation}The technical details are explained in Section
6.

In almost all studies of hard processes, where a mixture of
perturbative and non-perturbative QCD occurs, nucleons are taken
to be in helicity states moving with high energy along the
$z$-axis, and typically one is utilizing matrix elements of local
products of quark or gluon field operators between these states.
For these operators there is no problem in dealing with diagonal
matrix elements. But when it comes to an angular momentum sum rule
for the nucleon , care must be taken to decide whether one is
dealing with helicity states
 $| p_{z},\lambda\ra$, where
$p_{z} = (E,0, 0, p)$ or with canonical states $|p_{z},
\bm{s}_z\ra$, where $\bm{s}_z = (0, 0, 2\lambda)$.  The point is
that even though the initial states are the same,
\begin{equation}
| p_{z}, \lambda \ra = | {p_{z},\bm{s}_z}\ra \label{lpss}
\end{equation}
the singular nature of $J_i$ forces one to deal with non-diagonal
matrix elements i.e. to utilize $\la p' ,\sigma |$ where $\gb{p'}$
is not along the $z$-axis, and for these
\begin{equation}
\la p', \lambda | \neq \la p', \bm{s}_z| \label{apll}
\end{equation}

In this paper we show that it is possible to give a rigorous
derivation of the structure of the expectation values for
canonical states
\begin{equation}
\la p, \bm{s}| J_i | p, \bm{s}\ra \equiv \mathcal{L}_{p'
\rightarrow p} \la p',\bm{s} | J_i | p, \bm{s}\ra \label{pj22}
\end{equation}
where $\bm{s}$ is a unit vector along the rest frame spin
eigenvector, and for helicity states
\begin{equation}
\la p,\lambda |J_i| p, \lambda\ra \equiv \mathcal{L}_{p'
\rightarrow p} \la p', \lambda | J_i | p, \lambda \ra
\label{phlj}.
\end{equation}

 In general, for the arbitrary $i^{th}$ component of $\bm{J}$, for
  spin $ \frac{1}{2}$
 \begin{equation}
 \la p, \lambda |J_i|p,\lambda \ra \neq  \la p, \bm{s} = 2 \lambda \hat{\bm{p}}|J_i|p,\bm{s} = 2 \lambda \hat{\bm{p}} \ra
 \end{equation}
 even though $\bm{s}$ lies along the direction of $\hat{p}$ in both cases, and even if $\bm{p}$ is along OZ where Eq.(\ref{lpss}) holds.
 Only for the specific component of $\bm{J}$  along $\hat{\bm{p}}$ do the matrix elements agree, i.e. for arbitrary $\bm{p}$,
 \begin{equation}
 \la p, \lambda | \bm{J \cdot p}|p, \lambda\ra = \la p, \bm{s}= 2 \lambda \hat{\bm{p}}|\bm{J \cdot p} | p, \bm{s}= 2 \lambda \hat{\bm{p}}\ra.
 \end{equation}

 In using the sum rules
based on Eq.(\ref{pj22}) or Eq.(\ref{phlj}) for arbitrary $i$
  to test any model of the
nucleon in terms of its constituents, it is essential to construct
wave-functions appropriate to the type of spin state being used
for the nucleon. The equations Eq.(\ref{phlj}) and Eq.(\ref{pj22})
contain delta functions and {\em derivatives} of delta-functions
and this is the reason for the special care required. Throughout
this paper, with the exception of the discussion in Section~6 and
in Section~7, we will use canonical spin states. In these latter
sections we shall utilize Jacob-Wick helicity states for the
massless gluons.

In Section~4 we summarize the treatments of Jaffe-Manohar and
Shore-White. The approaches in Refs.~\cite{ref.03} and
\cite{ref.04} are different, so we shall present both in some
detail and will comment upon the dubious steps.

\section{The Jaffe-Manohar and Shore-White Approaches}

These authors employ the standard approach of trying to relate the
matrix elements of the angular momentum operators to those of the
energy-momentum density operator, and utilize the version
Eq.~(\ref{eq.05}) based on the symmetrized energy-momentum tensor.

\subsection{The Jaffe-Manohar treatment}
\label{sec.2.2}

In order to make efficient use of the Lorentz invariance the
authors of \cite{ref.03} prefer to label their states using the
covariant spin 4-vector $\mathcal{S}$ and to  consider the entire
tensor $J^{\mu\nu\lambda}$ and to integrate the tensor densities
over four-dimensional space time, i.e. they consider
\begin{equation}
   {\cal M}^{\mu\nu\lambda} (p,k,p',\mathcal{S}) \equiv \int d^4 x e^{i k\cdot x}
   \langle p', {\mathcal{S'}} = \mathcal{S} | M^{\mu\nu\lambda} (x) | p\, , \mathcal{S} \rangle
\label{eq.15}
\end{equation}
and eventually take the limit $k^\mu \rightarrow 0$. In
Ref.~\cite{ref.03} the LHS is written in the abbreviated form
${\cal M}^{\mu\nu\lambda} (p,k,s)$, but we shall use the above
notation for clarity. Note that Jaffe and Manohar use the notation
$s$ to mean the covariant spin which we denote by $\mathcal{S}$.
It is $2/ M$ times the expectation value of the Pauli-Lubanski
operator; see e.g. \cite{Leader}:
\begin{equation}
\mathcal{W}_{\mu}=- \half \epsilon_{\mu \nu \rho \sigma}P^{\nu}
J^{\rho \sigma} \label{P-L}
\end{equation}
and
\begin{equation}
(2 \pi)^3 p_0 M \delta^3(\bm{p}'  - \bm{p}) \mathcal{S}_{\mu}=\la
p', \bm{s} | \mathcal{W}_{\mu} |p,\bm{s} \ra \label{covariantS}
\end{equation}
In terms of components
\begin{eqnarray} \label{covcompS}
\mathcal{S}_0 & =& \frac{\gb{p \cdot s}}{M} \nonumber \\
\mathcal{S}^i  &=& s^i + p^i \frac{\gb{p \cdot s}}{M (p_0 + M)}
\end{eqnarray}
Also we take $\mathcal{S}^2 =-1$ while Ref.[4] takes the covariant
normalization to be $-M^2$.

The Lorentz invariant normalization of the states is conventional
\begin{equation}
   \langle p', \mathcal{S} | p, \mathcal{S} \rangle =
   (2\pi)^3 \, 2 p_0 \,\delta^3 (\bm{p}^{\,\prime} - \bm{p}).
\label{eq.16}
\end{equation}
The extra integration in Eq.~(\ref{eq.15}) $\int dt$, is argued to
be harmless, leading to an infinite $\delta(0)$ which cancels out
when calculating a genuine expectation value.

\noindent \underline{Comment 1} It will be seen in Section~5.3
that the choice $\mathcal{S'} = \mathcal{S}$ as done in
\cite{ref.03} is not consistent in a proper wave-packet treatment.

Analogously to the steps leading to Eq.~(\ref{ambiguous}) we have
\begin{eqnarray}
   {\cal M}^{\mu\nu\lambda} (p,k,p',\mathcal{S}) & = & \int d^4 x
e^{i {x\cdot (k-p+p')}}
   x^\nu \langle p',\mathcal{S} | T^{\mu\lambda} (0) | p,\mathcal{S} \rangle
   - (\nu \leftrightarrow \lambda)
   \nonumber \\
   & = & \int d^4 x  \left[ -i\frac{\partial}{\partial k_\nu}
e^{i {x\cdot (k-p+p')}}
   \right] \langle p',\mathcal{S} | T^{\mu\lambda} (0) | p,\mathcal{S} \rangle
   - (\nu \leftrightarrow \lambda)
   \nonumber \\
   & = &
   -i (2\pi)^4  \frac{\partial}{\partial k_\nu}  \left[ \delta^4 (k-p+p')
   \langle p', \mathcal{S}| T^{\mu\lambda} (0) | p,\mathcal{S} \rangle \right]
   - (\nu \leftrightarrow \lambda) .
\label{eq.17}
\end{eqnarray}

\noindent\underline{Comment 2} The last step in Eq.~(\ref{eq.17})
 can only be justified if $p'$ in
Eq.~(\ref{eq.15}) is considered an independent variable. We may
not take $p' = p-k$. Once this is recognized it is no longer so
evident that Eq.~(\ref{eq.15}) followed by the limit $k^\mu \to 0$
provides a natural definition of the ambiguous forward matrix
element.

Continuing from Eq.~(\ref{eq.17}) one has
\begin{eqnarray}
   {\cal M}^{\mu\nu\lambda} (p,k,p',\mathcal{S}) & = & -i (2\pi)^4 \left[
   \langle p-k,\mathcal{S} | T^{\mu\lambda} (0) | p,\mathcal{S} \rangle
   \frac{\partial}{\partial k_\nu} \delta^4(k-p+p') \right.
   \nonumber \\
   & & \left. + \delta^4(k-p+p') \frac{\partial}{\partial k_\nu}
   \langle p-k,\mathcal{S} | T^{\mu\lambda (0)} | p,\mathcal{S} \rangle \right]
   - (\nu \leftrightarrow \lambda) .
\label{eq.21}
\end{eqnarray}
In Ref.~\cite{ref.03} the limit $k^\mu \to 0$ is given as
\begin{eqnarray}
   {\cal M}^{\mu\nu\lambda} (p,0,p',\mathcal{S}) & = & -i (2\pi)^4  \bigg[
    \langle p,\mathcal{S}| T^{\mu\lambda} (0) | p,\mathcal{S} \rangle
   \partial^\nu \delta^4(0)   \nonumber \\
   & &  \left. + \delta^4(0) \frac{\partial}{\partial k_\nu}
   \langle p-k, \mathcal{S} | T^{\mu\lambda (0)} | p,\mathcal{S} \rangle  \right]
   - (\nu \leftrightarrow \lambda) .
\label{eq.22}
\end{eqnarray}
This form is a little puzzling, given that $p' \not= p-k$ in
Eq.~(\ref{eq.15}), as discussed in Comment~2. We thus prefer to
write Eq.~(\ref{eq.22}) in the form
\begin{eqnarray}
   {\cal M}^{\mu\nu\lambda} (p,0,p',\mathcal{S}) & = & -i (2\pi)^4 \left[
   \langle p-k, \mathcal{S}| T^{\mu\lambda} (0) | p,\mathcal{S} \rangle
\lim_{k^\mu \to 0} \frac{\partial}{\partial k_\nu}\delta^4(k-p+p')
\right.
   \nonumber \\
   & &\left.  + \delta^4(p'-p)
   \left.\frac{\partial}{\partial k_\nu}
   \langle p-k,\mathcal{S} | T^{\mu\lambda (0)} | p,\mathcal{S} \rangle
\right|_{k^\mu \to 0}
   \right]
   - (\nu \leftrightarrow \lambda) .
\label{eq.23}
\end{eqnarray}

The highly singular first term in Eq.~(\ref{eq.23}) can only be
understood in a wave-packet analysis. It corresponds to the
angular momentum about the origin arising from the motion of the
center of mass of the wave-packet, and has nothing to do with the
internal structure of the nucleon. Thus we will take as the
definition of the ambiguous forward matrix element in the
Jaffe-Manohar (JM) approach
\begin{eqnarray}
   {\cal M}^{\mu\nu\lambda} (p,0,p',\mathcal{S}) & = & -i (2\pi)^4
   \delta(0) \delta^3(\bm{p}^{\,\prime} -\bm{p})
   \left.\frac{\partial}{\partial k_\nu}
   \langle p-k, \mathcal{S} | T^{\mu\lambda }(0) | p,\mathcal{S} \rangle
\right|_{k^\mu \to 0}
   \nonumber \\
   & & - (\nu \leftrightarrow \lambda) ,
\label{eq.24}
\end{eqnarray}
where we have used the fact that for the on mass-shell momenta
$\bm{p}^{\,\prime} = \bm{p}$ forces $p'_0 = p_0$.

The last part of the analysis concerns the structure of the matrix
element of $T^{\mu\lambda} (0)$. For Eq.~(\ref{eq.24}) we require
an expansion in $k^\mu$ of  $\langle
p-k,\mathcal{S}|T^{\mu\lambda} (0) | p,\mathcal{S} \rangle$ up to
terms linear in $k^\mu$. It is at this point that the choice in
Eq.~(\ref{eq.15}) of $\mathcal{S}' = \mathcal{S}$ becomes
significant: it greatly simplifies the tensorial structure of the
expansion.

Following Ref.~\cite{ref.03} one writes, with $P^\mu = p^\mu +
\half k^\mu$,
\begin{equation}
   \langle p-k,\mathcal{S}| T^{\mu\lambda} (0) |p, \mathcal{S}\rangle =
   A_0 (k^2) P^\mu P^\lambda
   + i A_1(k^2) [ \epsilon^{\mu\alpha\beta\sigma} P^\lambda
                + \epsilon^{\lambda\alpha\beta\sigma} P^\mu ]
   k_\alpha P_\beta \,\mathcal{S}_\sigma + O(k^2),
\label{eq.25}
\end{equation}
where $A_0$ and $A_1$ are scalar functions of $k^2$. The terms in
Eq.~(\ref{eq.25}) are chosen to respect the relation

\begin{equation}
   k_\mu \langle  p-k, \mathcal{S} | T^{\mu\lambda} (0) |p, \mathcal{S} \rangle = 0.
\label{eq.26}
\end{equation}

(To see that $k \cdot P = 0$ one should recall that the nucleon
states are physical states with $(p-k)^2 = p^2 = M^2$.)

For the forward matrix element
\begin{equation}
   \langle p,\mathcal{S} | T^{\mu\nu} (0) | p,\mathcal{S}\rangle = A_0 (0)
p^\mu p^\lambda . \label{eq.27}
\end{equation}

\noindent\underline{Comment~3} While Eq.(\ref{eq.26}) \emph{is}
correct the crucial expansion (\ref{eq.25}) is not. The reason is
the following.The RHS of Eq.~(\ref{eq.25}) has been constructed to
transform under Lorentz transformations as a genuine second-rank
tensor on the grounds that $T^{\mu\lambda} (0)$ transforms as a
second-rank tensor. But the \emph{non-forward} matrix elements of
a tensor operator do \emph{not} transform covariantly, as was
explained in the discussion of Eq.~(\ref{tensorform1}). We shall
see the consequences of this in Section~5.1.

Continuing with the derivation in Ref.~\cite{ref.03}, we have from
Eqs.~(\ref{eq.08}) and (\ref{eq.16})
\begin{eqnarray}
   \langle p', \mathcal{S}| \int d^3 x T^{00} (x) | p,\mathcal{S} \rangle & = &
   \langle p',\mathcal{S}|  H | p,\mathcal{S}\rangle  \nonumber \\
   & = & 2p_0^2 (2\pi)^3 \delta^3 (\bm{p}^{\,\prime} - \bm{p}),
\label{eq.28}
\end{eqnarray}
where we have used the fact that $P^0$ is the total energy or
Hamiltonian operator.
 But the LHS,
%taking $t=0$
%%BB
taking $t=0$, equals
%%BB
%
%\begin{eqnarray}
%\label{eq.29}
%   \rule{10ex}{0ex} & = & \int d^3 x
%   e^{i \bm{x \cdot (p-p')}}
%   \langle p',\mathcal{S} | T^{00} (0) | p, \mathcal{S} \rangle
%   \nonumber \\
%   & = & (2\pi)^3 \delta^3 (\bm{p}^{\,\prime} - \bm{p}) A_0(0)\, p_0^2
%\end{eqnarray}
%
\begin{equation}
\label{eq.29}
   \int d^3 x e^{i \bm{x \cdot (p-p')}}
   \langle p',\mathcal{S} | T^{00} (0) | p, \mathcal{S} \rangle
    = (2\pi)^3 \delta^3 (\bm{p}^{\,\prime} - \bm{p}) A_0(0)\, p_0^2
\end{equation}
from Eq.~(\ref{eq.27}).

Comparing Eqs.~(\ref{eq.28}) and (\ref{eq.29}) yields
\begin{equation}
   A_0 (0) = 2.
\label{eq.30}
\end{equation}
Now one uses Eq.~(\ref{eq.25}) to calculate the derivative needed
in Eq.~(\ref{eq.24}). The result in Ref.~\cite{ref.03} is
\begin{equation}
   {\cal M}^{\mu\nu\lambda} (p,0,p',\mathcal{S}) = (2\pi)^4 \delta(0)
   \delta^3(\bm{p}^{\,\prime} - \bm{p}) A_1 (0)
   \left[
   2 p^\mu \epsilon^{\lambda\nu\beta\sigma}
   - p^\nu \epsilon^{\mu\lambda\beta\sigma}
   +  p^\lambda \epsilon^{\mu\nu\beta\sigma} \right] p_\beta \mathcal{S}_\sigma.
\label{eq.31}
\end{equation}

 Lastly the value of $A_1 (0)$ is
found by choosing a nucleon state at rest and spin along $OZ$ This
is an eigenstate of $J_z$
\begin{equation}
 J_z | 0, \hat{\bm{z}}\ra = \half |0, \hat{\bm{z}}\ra
\label{eq.32}
\end{equation}
Then with $\Sc = (0,0,0,1)$
\begin{eqnarray}
   \langle p', \hat{\bm{z}} | \int d^4 x M^{012} (x) | 0, \hat{\bm{z}} \rangle & = &
   \int dt \langle p', \Sc | J_z | 0, \Sc \rangle \nonumber \\
   & = & \half \, (2\pi)^4 2M \delta^3 (\bm{p}') \delta(0).
\label{eq.33}
\end{eqnarray}
But from Eq.~(\ref{eq.31}) the LHS is just equal to
\footnotemark[2] \footnotetext[2]{There is a factor of $M^2$
missing in the value of $A_1(0)$ given in \cite{ref.03}.}
\begin{equation}
{\cal M}^{012} (0, 0, p', \Sc ) = (2\pi)^4 \delta(0)\, \delta^3
   (\bm{p}') A_1 (0) \,2M^2
\label{eq.34}
\end{equation}
so
\begin{equation}
   A_1 (0) = \frac{1}{2M} .
\label{eq.35}
\end{equation}

Finally, then, in the Jaffe-Manohar treatment the interpretation
given to the forward matrix element of the angular momentum
operator is
\begin{eqnarray}
   \langle p',\mathcal{S} | \int d^4x M^{\mu\nu\lambda} (x) | p,\mathcal{S}\rangle & = &
   {\cal M}^{\mu\nu\lambda} (p,0,p', \mathcal{S}) \nonumber \\
   & = & \frac{(2\pi)^4 \delta(0) \delta^3 (\bm{p}^{\,\prime} - \bm{p})}{2M}
   \left[ 2 p^\mu \epsilon^{\lambda\nu\beta\sigma} -
   p^\nu \epsilon^{\mu\lambda\beta\sigma} +
   p^\lambda \epsilon^{\mu\nu\beta\sigma} \right] p_\beta \mathcal{S}_\sigma .
   \nonumber \\
\label{eq.36}
\end{eqnarray}

Equation (\ref{eq.36}) is meant to provide a general basis for
angular momentum sum rules. So, for example, if we have a theory
of the nucleon in terms of quark and gluon fields and we construct
the operator $M^{\mu\nu\lambda}$ from these fields, then the
requirement that our $M^{\mu\nu\lambda}$ satisfy Eq.~(\ref{eq.36})
for an arbitrary state of the nucleon yields a set of conditions
on some of the elements of the theory.

As indicated in the comments, there are flaws in the derivation
and Eq.~(\ref{eq.36}) is incorrect.  We shall present the correct
result, in a wave-packet treatment in Sections~5.1 and 5.2. There
our states or wave-functions will be normalized to $1$ so that we
calculate actual expectation values. Moreover, as will be
explained in Section~4.2, the wave-packet approach seems only able
to deal with the physically relevant operators $\int d^3x M^{0ij}
(x)$. For these conserved densities the $\int dt$ in
Eq.~(\ref{eq.15}) is simply equivalent to the factor $2\pi
\delta(0)$ in Eq.~(\ref{eq.31}). Thus, dividing the expression
(\ref{eq.36}) by this factor and by the normalization given in
Eq.~(\ref{eq.16}) we have, for the expectation values in the
JM-treatment
\begin{equation}
   \left. \frac{\langle p,\mathcal{S} | \int d^3x M^{0ij} (x) | p,\mathcal{S}
\rangle}{\langle p,\mathcal{S}| p,\mathcal{S} \rangle}
\right|_{\rm JM} = \frac{1}{4M p_{0}} \left[
   2 p^0 \epsilon^{ji\beta\sigma}
   - p^i \epsilon^{0j\beta\sigma} + p^j \epsilon^{0i\beta\sigma} \right]
   p_\beta \mathcal{S}_\sigma
\label{eq.37}
\end{equation}
which can be compared directly with the result we shall obtain in
Section~5.1 and Section~5.2. In Section~6 we shall give a
completely independent corroboration of these results.

We turn now to the treatment of Shore and White, which basically
follows the pattern of Ref.~\cite{ref.03}, but attempts to put the
argument on a rigorous footing via the use of wave-packets.

\subsection{The Shore-White treatment} \label{sec.2.3}

As already mentioned, the authors of Ref.~\cite{ref.03} remark
that a wave-packet approach is needed to justify some of the
manipulations involved, more precisely, to get rid of the
unwelcome derivatives of $\delta$-functions. This is done in
Ref.~\cite{ref.04}, but, as we shall see, the treatment  also
suffers from some of the incorrect elements commented on in
Section~4.1. The notation in Ref.~\cite{ref.04} differs somewhat
from that of Ref.~\cite{ref.03}, so we will rephrase the notation
in Ref.~\cite{ref.04} to match as closely as possible the
development in Ref.~\cite{ref.03}.

The authors of Ref.~\cite{ref.04} try to give a sensible
definition to the forward matrix elements defined in
Eq.~(\ref{ambiguous}), i.e.

\begin{equation}
   {\cal M}^{0ij}(p, \mathcal{S}) = \langle p,\mathcal{S} | \int d^3x
   \left[x^i T^{0j} (x) - x^j T^{0i} (x) \right] | p,\mathcal{S} \rangle
\label{eq.38}
\end{equation}
by utilizing a wave packet. Thus they define
\begin{equation}
   | \phi (p,\mathcal{S}) \rangle = \int \frac{d^3q}{(2\pi)^3 \sqrt{2 q_0}}
   \phi[(\bm{q} - \bm{p})^2] | q,\mathcal{S} \rangle
\label{eq.39}
\end{equation}
where $\phi$ drops rapidly to zero as $|\bm{q} - \bm{p}| \to
\infty$, and they interpret Eq.~(\ref{eq.38}), modulo
normalization, as
\begin{equation}
   {\cal M}^{0ij}_{\rm SW}(p,\mathcal{S}) = \langle \phi(p,\mathcal{S}) | \int d^3x
   \left[x^i T^{0j} (x) - x^j T^{0i} (x) \right] | p,\mathcal{S} \rangle .
\label{eq.40}
\end{equation}
It turns out to be sufficient to use just one wave packet, either
for the inital or the final state.

Note that this differs from ${\cal M}^{\mu\nu\lambda}
(p,0,p',\mathcal{S})$ in two respects. Firstly, consideration is
given here only to the spatial elements $i,\, j$ of the tensor and
secondly, the integral is over three-dimensional space. The latter
difference is not significant given that the operators in
Eq.~(\ref{eq.38}) are supposed to be time-independent.

The derivation then runs as follows:
\begin{eqnarray}
   {\cal M}_{\rm SW}^{0ij} (p,\mathcal{S}) & = & \int \frac{d^3q}{(2\pi)^3}
   \phi[(\bm{q} - \bm{p})^2]
   \left[ x^i e^{i \bm{x \cdot (p-q)}}
   \langle q,\mathcal{S} | T^{0j} (0) | p,\mathcal{S} \rangle -(i
\leftrightarrow j) \right] \nonumber \\
   & = & i \int \frac{d^3q}{(2\pi)^3}  \phi[(\bm{q} - \bm{p})^2]
   \int d^3x
   \left( \frac{\partial}{\partial q_i}
   e^{i \bm{x \cdot (p-q)}} \right)
   \langle q,\mathcal{S} | T^{0j} (0) | p,\mathcal{S}\rangle \nonumber  \\
 & & -(i \leftrightarrow j) . \label{eq.42}
\end{eqnarray}
Exchanging the order of integration, then integrating by parts
w.r.t. $\bm{q}$, and discarding the surface terms at $q_i = \pm
\infty$,
\begin{eqnarray}
   {\cal M}_{\rm SW}^{0ij} (p,\mathcal{S}) & = & -i \int d^3 x \int
\frac{d^3q}{(2\pi)^3}
   e^{i\bm{x \cdot (q-p)}}
    \frac{\partial}{\partial q_i}  \left[
 \phi \left( ( \bm{q} - \bm{p})^2 \right)
   \langle q,\mathcal{S} | T^{0j} (0) | p,\mathcal{S} \rangle  \right]
   \nonumber \\
   & = & - i \int d^3q \, \delta^3(\bm{q} - \bm{p})
   \left[
   \left(\frac{\partial \phi}{\partial q_i} \right)  \langle
   q,\mathcal{S}| T^{0j} (0) | p,\mathcal{S} \rangle +
   \phi \frac{\partial}{\partial q_i} \langle q,\mathcal{S }| T^{0j} (0)
   |p,\mathcal{S} \rangle
   \right]
   \nonumber \\
   & = & - i
   \left.
   \frac{\partial \phi[(\bm{q} - \bm{p})^2]}{\partial q_i}
   \right|_{\bm{q} = \bm{p}}
   \langle p,\mathcal{S} | T^{0j} (0) | p,\mathcal{S }\rangle +
   \phi (0)
   \left.
   \frac{\partial}{\partial q_i} \langle q,\mathcal{S} | T^{0j} (0) |
p,\mathcal{S} \rangle
   \right|_{\bm{q} = \bm{p}}
    ,
   \nonumber \\
\label{eq.45}
\end{eqnarray}
all antisymmetrized under $i \leftrightarrow j$.

Now
\begin{equation}
  \left. \frac{\partial \phi[(\bm{q} - \bm{p})^2]}{\partial q_i}
\right|_{\bm{q} = \bm{p}}=  0, \label{eq.46}
\end{equation}
so in the Shore-White (SW) approach
\begin{equation}
   {\cal M}^{0ij}_{\rm SW} (p,\mathcal{S})  =
   -i \phi(0) \left[ \left.
   \frac{\partial}{\partial q_i} \langle q,\mathcal{S }| T^{0j} (0) |p,\mathcal{S} \rangle
   \right|_{\bm{q} = \bm{p}}   -(i \leftrightarrow j)
   \right] ,
\label{eq.47}
\end{equation}
which is identical to ${\cal M}^{0ij}_{\rm JM}
(p,0,p',\mathcal{S})$ in Eq.~(\ref{eq.24}), aside from
normalization, which is irrelevant, since, at the end, it cancels
out when computing actual expectation values. \\

\noindent\underline{Comment~4} It is actually {\em not} possible
in Eq.~(\ref{eq.47}) to take the same $\mathcal{S}$ in both
initial and final states,  as in eq.~(\ref{eq.15}). The reason is
the following. A wave packet is, by definition, a superposition of
{\em physical} states. But for a spin-1/2 particle in a physical
state

\begin{equation}
   q^2 = M^2 \quad \mbox{\underline{and}}\quad q \cdot \mathcal{S}= 0.
\label{eq.48}
\end{equation}
and we cannot integrate independently over each component of
$\bm{q}$. That is one reason why we have chosen to define the
states using the rest-frame spin vector $\bm{s}$ instead.
The correct procedure is then to take the same rest frame $\bm{s}$
for the initial and final states, not the same $\mathcal{S}$. Thus
the wave packet Eq.~(\ref{eq.39}) should be modified to
\begin{equation}
   | \phi (p, \bm{s}) \rangle = \int \frac{d^3 q}{(2\pi)^3}
   \phi[(\bm{q} - \bm{p})^2] | q, \gb{s} \rangle
\label{eq.50}
\end{equation}
and all three components of $\bm{q}$ can then be integrated over
independently.
%%%%%%%%%%%%%%%%%%%%%%%%
%Since the expansion Eq.~(\ref{eq.25}) is in any case not correct
%in general, because, as explained in Comment~3, it is based upon
%an incorrect application of Lorentz covariance, we shall not
%pursue the matter further here.
%%%%%%%%%%%%%%%%%%%%%%%%%%%%

The use of the Lorentz tensor form, as in Eq.~(\ref{eq.36}), is
central to some of the most interesting arguments given by Shore
and White---for example their conclusion that the axial charge
does not contribute to the angular momentum sum rules. But, as
pointed out in Comment 3, this form is based on the incorrect
expansion Eq.~(\ref{eq.25}), and so one must be skeptical about
their results. We shall see, however, in Section~5.2, that the
conclusion regarding the role of the axial charge in the sum rules
\textit{is} correct in spite of the erroneous argument.

 To summarize: the Shore-White wave-packet analysis
provides a justification for the manipulations in the
Jaffe-Manohar treatment for the spatial components of the angular
momentum tensor. But the analysis is not correct in general
because it utilizes the impermissible simplification
$\mathcal{S}=\mathcal{S}'$ and also makes an incorrect use of
Lorentz covariance.

Given these criticisms, it is interesting to consider one
important practical case of a sum rule that can be derived from
Eq.~(\ref{eq.36}).  Namely, for a longitudinally polarized proton
moving along $OZ$ i.e. in a state of definite helicity $\half$
with $\gb{p}$ along $OZ$ one has the sum rule
\begin{equation}
\half = \half \Delta \Sigma + \Delta G + \langle L^q \rangle +
\langle L^G \rangle \label{eq.57}
\end{equation}
where $\Delta\Sigma$,$\Delta G$ are the first moments of the
polarized quark flavour-singlet and the polarized gluon densities,
respectively, and $\langle L^q \rangle $ , $ \langle L^G \rangle $
are contributions from the quark and gluon internal orbital
momenta. This sum rule is so manifestly intuitively right that it
is almost impossible to contemplate it being incorrect \cite{Ji}.
And indeed, it is correct.  This is the one unique case where the
flaws in the deduction are irrelevant, as will be seen later.

In Section~\ref{sec.3} we will present what we believe to be a
correct evaluation of the wave-packet definition of the
recalcitrant forward matrix element.

\section{A detailed wave-packet treatment}
\label{sec.3}

Given our critical stance {\em vis-\`{a}-vis} the treatments of
Refs.~\cite{ref.03} and \cite{ref.04} it is incumbent upon us to
proceed with caution. We shall therefore present a detailed
wave-packet treatment for two separate situations, for a
relativistic quantum-mechanical Dirac particle and for the
field-theoretical case. It will emerge that there is complete
agreement between the results for the two cases and that they
differ, in general, from Eq.~(\ref{eq.37}).

We shall also address the question as to the validity of
Eq.~(\ref{eq.2.12}), which was assumed in the derivations in Refs.
\cite{ref.03} and \cite{ref.04}. In order to do this we shall need
to distinguish between $M^{\mu\nu\lambda}_{\rm orb} (x)$
constructed out of the canonical energy-momentum density
$T^{\mu\lambda}_{\rm C} (x)$ as in Eq.~(\ref{eq.04}) and
$M^{\mu\nu\lambda} (x)$ used in Section~4.1 and Section~4.2 based
upon the symmetrized $T^{\mu\lambda} (x)$, as in
Eq.~(\ref{eq.05}).

\subsection{Relativistic quantum-mechanical Dirac particle}
\label{sec.3.1}

We construct the wave-function corresponding to a superposition of
physical states centered around momentum $\bm{p}$, all of which
have rest-frame spin vector $\bm{s}$:

\begin{equation}
   \psi_{p,\bm{s}} (\bm{x},t) = \frac{\bar{N}}{(2\pi)^3}
   \int \frac{d^3 q}{q^0} e^{- \lambda^2 \bm{(q-p)}^2}
   e^{i (\bm{q \cdot x} - q^0 t)}
   u(q, \bm{s}), \label{eq.48a}
\end{equation}
where $q^0 = \sqrt{\bm{q}^2 + M^2}$, and
\begin{equation}
   u(q, \bm{s}) = \sqrt{\frac{q^0 +M}{2 M}}
   \left(
   \begin{array}{c}
1 \\ \frac{\displaystyle{\bm{\sigma \cdot q}}}{\displaystyle{q^{0}
+ M}}
   \end{array}
   \right) \chi(\bm{s})
\label{eq.49}
\end{equation}
with
\begin{equation}
   \chi^\dagger(\bm{s}) \gb{\sigma} \chi(\bm{s}) = \bm{s}.
\label{eq.51}
\end{equation}

The constant $\bar{N}$, whose value is irrelevant for the moment,
is adjusted so that $\psi_{\bm{p}, \bm{s}}$ is normalized to 1,
i.e.
\begin{equation}
\int d^3 x \psi^\dagger_{p,\bm{s}} (x) \psi_{p,\bm{s}} (x) = 1.
\label{eq.52}
\end{equation}
Since our aim is to provide a sensible prescription for a
plane-wave state of definite momentum $\bm{p}$, we shall, at the
end, take the limit $\lambda \to \infty$. In this limit the values
of $\bm{q}$ that contribute in the integral Eq.~(\ref{eq.48a}) are
forced towards $\bm{p}$, so we make a Taylor expansion of
$(1/q^0)\,u(q,\bm{s})$ about the point $\bm{p}$ and keep only
those terms that survive ultimately in the limit $\lambda \to
\infty$.

Note that we are unable to carry out the analysis if $t$ can be
arbitrarily large. The point of using the wave-packet is to
produce a cut-off in the divergent spatial integrals in
Eq.~(\ref{ambiguous}), but this does not produce a cutoff in $t$ .
Thus the wave-packet approach can not be used if the operator
densities are integrated over four-dimensional space-time. Since
we are only interested in conserved densities, we will take
advantage of the time-independence to choose $t = 0$ in
Eq.~(\ref{eq.48a}).

It turns out to be sufficient to keep just the first two terms of
the Taylor expansion, which yield:
\begin{eqnarray}
   \frac{1}{q^0} u(q, \bm{s}) & = &
   \frac{1}{p^0}\ \left\{ \rule{0mm}{10mm} u(p, \bm{s}) -
    \frac{1}{p^0 + M}
   \Bigg[
   \frac{p^0 + 2M}{2p^{0\,2}} u(p, \bm{s})  \bm{p} +  \right.
   \nonumber \\
   && \left. \left. \sqrt{\frac{p^0 + M}{2M}}
  \left(
   \begin{array}{c}
   0 \\
\frac{\displaystyle{\bm{\sigma} \cdot \bm{p}\,
\bm{p}}}{\displaystyle{p^0 (p^0 + M)}} - \bm{\sigma}
   \end{array}
   \right) \chi(\bm{s})
   \right]
   \cdot (\bm{q} - \bm{p})
   \right\}
\label{eq.53}
\end{eqnarray}
with $p^0 = \sqrt{\bm{p}^2 + M^2}$.

The term $(\bm{q-p})$ is removed from under the integral in
Eq.~(\ref{eq.48a}) via
\begin{equation}
   (\bm{q-p})e^{-\lambda^2(\bm{q-p})^2} =
   \frac{\bm{\nabla}_{\bm{p}}}{2\lambda^2} e^{-\lambda^2
(\bm{q-p})^2}. \label{eq.54}
\end{equation}

The remaining integral is just the Fourier transform of a
Gaussian, yielding a factor $\exp(-\bm{x}^{\, 2}/(4\lambda^2)) \,
\exp(i\bm{p}\cdot \bm{x})$, and the $\bm{\nabla}_{\bm{p}}$ on the
RHS of Eq.~(\ref{eq.54}) then produces a term $i\bm{x}$. The
result is
\begin{equation}
   \psi_{p,\bm{s}} (\bm{x}, 0) = N e^{-\bm{x}^2/4 \lambda^2}
   e^{i \bm{p \cdot x}} u_{p,\bm{s}} (x)
\label{eq.55}
\end{equation}
where
\begin{equation}
   N^2 = \frac{M}{p^0} \left( \frac{1}{\sqrt{2\pi} \lambda} \right)^3
\label{eq.56}
\end{equation}
and
\begin{eqnarray}
   u_{p,\bm{s}} (x) &  = &
   \left[ 1 - \frac{i \bm{x \cdot p}}{2\lambda^2}
   \frac{p^0 + 2M}{2p^{0\, 2} (p^0 + M)} \right] u(p,\bm{s}) + \nonumber \\
& &  + \frac{i}{2\lambda^2 (p^0 + M)} \sqrt{\frac{p^0 + M}{2M}}
\left(
   \begin{array}{c} 0 \\
   \bm{\sigma}\cdot\bm{x}
   - \frac{\displaystyle{\bm{x}\cdot \bm{p}\, \bm{\sigma} \cdot\bm{p}}}{\displaystyle{p^0 (p^0 + M)}}
\end{array} \right) \chi(\bm{s}) .
\label{eq.57a}
\end{eqnarray}

The structure of Eq.~(\ref{eq.57a}) is extremely instructive in
understanding both the difference between matrix elements of local
operators like $T^{\mu\lambda} (x)$ and compound ones like
$x^{\rho}\,T^{\mu\lambda} (x)$, and the question as to how many
terms of the Taylor expansion are necessary.

On the one hand the Gaussian implies that the effective values of
$\bm{x}$ satisfy $|\bm{x}| \leq 2 \lambda$. On the other, the
$n^{\rm th}$ term in the Taylor expansion provides a term of order
$(|\bm{x}|/\lambda^2)^n$. For a local operator there are no other
factors of $x$ present, so even the first-order term of a Taylor
expansion can be ignored in the limit $\lambda \to \infty$. For
the angular momentum, on the contrary, there is one explicit
factor of $x$. The first-order term in the Taylor expansion is
then essential, but higher order terms can be disregarded as
$\lambda \to \infty$. Of course, this is totally analogous to what
happened in Section~4.1 , where the results involved a first
derivative. The difference is that here the derivative has been
calculated accurately and without recourse to an incorrect
assumption of Lorentz covariance.

Now that a sufficiently accurate wave function has been obtained,
we turn to the calculation of the expectation values of the
operators. In Dirac theory the canonical energy-momentum density
is \cite{JR}
\begin{equation}
   T^{\mu\lambda}_{\rm C} (x) = \frac{i}{2}
   \bar{\psi} (x) \gamma^\mu \partial^\lambda \psi (x) + \;{\rm h.c.}
\label{eq.58}
\end{equation}
and the orbital angular momentum density is
\begin{equation}
   M^{\mu\nu\lambda}_{\rm orb} (x) = x^\nu T^{\mu\lambda}_{\rm C} (x)
   -  x^\mu T^{\nu\lambda}_{\rm C} (x)
\label{eq.59}
\end{equation}
and the spin density is
\begin{equation}
   M^{\mu\nu\lambda}_{\rm spin} (x) = \half \bar{\psi} (x) \gamma^\mu
   \sigma^{\nu\lambda} \psi (x) .
\label{eq.60}
\end{equation}
The orbital angular momentum operators $M^{ij}_{\rm orb}$ and the
spin angular momentum $M^{ij}_{\rm spin}$ are the space integrals
of $M^{0ij}_{\rm orb}$ and $M^{0ij}_{\rm spin}$ respectively,
calculated, in our case, at $t = 0$ and are not separately time
independent.

The calculation of the spatial derivatives of the wave function
(\ref{eq.55}) needed in Eq.~(\ref{eq.58}) and the subsequent Dirac
algebra is straightforward, but laborious and will not be spelled
out here. Helpful is the fact that odd functions of $\bm{x}$ will
vanish under integration $\int d^3 x$ since the non-symmetric term
$\exp(i\gb{p \cdot x})$ cancels out in constructing
Eq.~(\ref{eq.58}). The terms in the spatial derivative that
survive give
\begin{equation}
   \partial^j \psi_{p,\bm{s}} (\bm{x}, 0) =
   iN e^{-\bm{x}^{\, 2}/4\lambda^2} e^{i\bm{p \cdot x}}
   u^j_{p,\bm{s}} (\bm{x})
\label{eq.61}
\end{equation}
where
\begin{eqnarray}
   u^j_{p,\bm{s}} (\bm{x}) & = &
   \left\{
   p^j \left[
    1- \frac{i(p^0 +2M)}{2\lambda^2 (p^0 +M)}
   \frac{\bm{p \cdot x}}{2p^{0\,2}}
   \right]
   + \frac{ix^j}{2\lambda^2}
   \right\} u(p,\bm{s}) \nonumber \\
  & & + \frac{ip^j}{2\lambda^2 (p^0 +M)} \sqrt{\frac{p^0 +M}{2 M}}
   \left(
   \begin{array}{c} 0 \\
   \bm{\sigma} \cdot \bm{x} -
   \frac{\displaystyle{\bm{p}\cdot\bm{x}\, \bm{\sigma}\cdot\bm{p}}}{\displaystyle{p^0 (p^0 + M)}}
   \end{array}
   \right) \chi(\bm{s}).
\label{eq.62}
\end{eqnarray}
Then, keeping only the terms that survive in the limit
$\lambda\to\infty$, we find ultimately:
\begin{equation}
   M^{ij}_{\rm orb} = \frac{1}{2p^0 (p^0 + M)}
   \left[ p^j (\bm{p}\times\bm{s})^i - p^i(\bm{p}\times\bm{s})^j\right]
\label{eq.63}
\end{equation}
or via the anologue of Eq.~(\ref{eq.09}), we express the orbital
angular momentum vector $\bm{L}$ in terms of the independent
vectors $\bm{p}$ and $\bm{s}$:
\begin{eqnarray}
   \bm{L} & = & -\frac{1}{2p^0 (p^0 + M)} \;
   [\bm{p}\times(\bm{p}\times\bm{s})] \nonumber \\
   & = &  \frac{1}{2p^0 (p^0+ M)} \; [\bm{p}^2 \bm{s} - (\bm{p} \cdot
\bm{s})\bm{p}]   \label{eq.64}
\end{eqnarray}
For the spin angular momentum we find
\begin{eqnarray}
   M^{ij}_{\rm spin} & = & \frac{\epsilon^{ijk}}{2p^0}
   \left[ M s^k + \frac{(\bm{p}\cdot\bm{s})p^k}{p^0 +M} \right]
   \nonumber \\
   & = & \frac{\epsilon^{ijk}}{2p^0} \mathcal{S}^k M
\label{eq.65}
\end{eqnarray}
where we have used  Eq.~(\ref{covcompS}).

For the spin vector $\bm{S}$ of the system Eq.~(\ref{eq.65})
yields
\begin{equation}
\bm{S} = \frac{1}{2p^{0}}\; \left[M\bm{s} + \frac{(\bm{p} \cdot
\bm{s}) }{p^{0} + M}\bm{p}\right]   \label{eq.69}
\end{equation}
Adding Eqs~(\ref{eq.64}) and (\ref{eq.69}) we have the remarkable
result that the term proportional to $\bm{p}$ cancels out, and
\begin{equation}
\label{eq.70} \bm{J} = \half \bm{s}.
\end{equation}

For later use we write this as
\begin{equation}
   J^{ij} = M^{ij}_{\rm orb} + M^{ij}_{\rm spin} = \half\epsilon^{ijk}s^k.
\label{eq.71}
\end{equation}

In the above we have used the canonical form of the total angular
momentum, where it is split naturally into an orbital and spin
part. In Section~4, however, following Refs~\cite{ref.03},
\cite{ref.04} we utilized the angular momentum tensor built from
the symmetrized Belinfante energy-momentum density. According to
Eq.~(\ref{eq.2.12}) this should yield the same forward matrix
element as the total angular momentum in the canonical approach.
For our quantum-mechanical example we can check whether indeed
$M^{ij} = J^{ij}$.

For the symmetrical energy-momentum density we have
$T^{\mu\lambda} = \half (T^{\mu\lambda}_{\rm C} +
T^{\lambda\mu}_{\rm C})$ so that from Eqs.~(\ref{eq.05}) and
(\ref{eq.58})
\begin{equation}
   M^{\mu\nu\lambda} (x) = \half  M^{\mu\nu\lambda}_{\rm orb} (x) +
   \half \left\{ \frac{i}{2} \left[
   x^\nu \bar{\psi} (x) \gamma^\lambda \partial^\mu \psi (x)  -
   x^\lambda \bar{\psi} (x) \gamma^\nu \partial^\mu \psi (x)
   \right] + \; {\rm h. c.} \right\}.
\label{eq.72}
\end{equation}
For $M^{0ij} (x)$ we now require the time-derivative $\partial
\psi_{p,\bm{s}} (\bm{x},t)/\partial t |_{t = 0}$ of the wave
function (\ref{eq.48a}). The time derivative gives a factor
$-iq^0$ which is expanded as
\begin{equation}
   -iq_0 = -i \left[ p_0 + \frac{\bm{p}\cdot(\bm{q} - \bm{p})}{p_0} \right]
\label{eq.73}
\end{equation}
and via the steps explained after Eq.~(\ref{eq.53}) this becomes a
multiplicative factor:
\begin{equation}
   \left. \frac{\partial}{\partial t} \psi_{p,\bm{s}} (\bm{x}, t)
   \right|_{t = 0} = -i \left[ p_0 + \frac{i\bm{p}\cdot\bm{x}}{2 \lambda^2 p_0}
   \right] \psi_{p,\bm{s}} (\bm{x}, 0) .
\label{eq.74}
\end{equation}
After further laborious Dirac algebra we find
\begin{equation}
   M^{ij} = \half \epsilon^{ijk}s^k
\label{eq.75}
\end{equation}
in complete agreement with Eq.~(\ref{eq.71}).

Thus despite the fact that $M^{0ij} (x)$ and $J^{0ij} (x)$ differ
by the divergence of a {\it compound} operator, and despite the
fact that the definition of the forward matrix element involves
non-forward ones, it seems that Eq.~(\ref{eq.2.12}) is indeed
valid, in a wave-packet approach, as was claimed in
Ref.~\cite{ref.04}.

In the next section we shall see that we obtain the same, to us
surprising, result Eq.(\ref{eq.71}) in field theory, and we shall
then compare it to the results in Sections~4.1 and 4.2.

\subsection{Field theoretic treatment}
\label{sec.3.2}

Analogously to the quantum mechanical case we construct a
wave-packet state $| \Psi_{p,\bm{s}} \ra$ as a linear
superposition of physical (canonical spin) plane wave states of
momentum $q$ all of which have the same rest-frame spin vector
$\bm{s}$:
\begin{equation}
| \Psi_{p,\bm{s}} \ra = \frac{N}{\sqrt{ (2\pi^3)} }\int d^{3} {q}
e^{-\lambda ^2(\bm{q- p)}^2} |q,\bm{s}\ra  \label{eq.3.29}
\end{equation}
Note that we label the momentum eigenstates by $\bm{s}$, not by
the covariant spin vector $\mathcal{S}$.  As before we shall
consider the limit $\lambda \rightarrow \infty$ and we normalize
the state to 1.
\begin{equation}
\la\Psi_{p,\bm{s}} | \Psi_{p,\bm{s}}\ra = 1 \label{eq.3.30}
\end{equation}
It follows that
\begin{equation}
N^2 = \left( \frac{2\lambda}{\sqrt{ 2\pi}} \right) ^3
\frac{1}{2p_0} \label{eq.3.31}
\end{equation}
with $p_0 = \sqrt{p^2 + M^2}$.

  Let us first consider the general
structure of the matrix elements of some local operator density
$O(x)$ between the states $|\Psi_{p,\bm{s}}\ra$. We have
\begin{equation}
\la\Psi_{p',\bm{s}}|O (\bm{x},0)|\Psi_{p,\bm{s}}\ra =
\frac{N^{2}}{(2\pi)^3} \int d^{3}q d^{3}q ' e ^{- \lambda^ 2
(\bm{p-q})^2} e ^{-\lambda^2 (\bm{p'-q'})^2} e^{i \bm{x \cdot
(q-q')}} \la q', \bm{s} | O(0) | q, \bm{s}\ra \label{eq.3.32}
\end{equation}
where we have temporarily kept $p'$ distinct from $p$ as an aid to
calculation.

The matrix element on the RHS is expanded in a Taylor series in
$q$ and $q'$ about the points $p$ and $p'$ respectively and to the
required order is of the form
\begin{equation}
\la q', \bm{s} | O(0) | q, \bm{s}\ra = f_{1} (p, \bm{s}) +\gb{
(q-p) }\cdot \bm{f _2} (p,\bm{s}) + \gb{(q' - p' )}\cdot \bm{f _3
} (p',\bm{s}) \label{eq.3.33}
\end{equation}

Factors like $\gb{q-p}$ are transformed into
$\frac{\bm{\nabla}_{p}}{2 \lambda ^2}$ etc. under the integral in
Eq.~(\ref{eq.3.32}), so that
\begin{equation}
\la\Psi_{ p',\bm{s} }| O (\bm{x},0) | \Psi_{p,\bm{s}} \ra = \left[
f_1 + \bm{f_2} \cdot \frac {\nabla_p}{2 \lambda ^2} + \bm{f_3}
\cdot \frac{\nabla_p'}{2 \lambda ^2}\right]
  N^2  \int d^3q  d^3q' e^{-\lambda^ 2 \bm{(q-p)}^2}e^{- \lambda
^2 \bm{(q'- p')}^2} e^{i\bm{x} \cdot \bm{(q - q')}}
\label{eq.3.34}
\end{equation}

Putting $\bm{r }= \bm{q-p}, \, \bm{ r'} =\bm{ q'- p'}$ the
integral in Eq.~(\ref{eq.3.34}) becomes
\begin{equation}
e^{i \bm{x \cdot (p-p' })}\int d^3 r e^{-\lambda^ 2 r^2} e^{i\bm{r
\cdot x}}\int d^{3} r' e^{-\lambda^ 2 r'^2} e^{-i \bm{r' \cdot x}}
\label{eq.3.36}
\end{equation}
so that $\nabla_p$ can be replaced by $i\bm{x}$ and $\nabla_{p' }$
by $-i\bm{x} $ in Eq.~(\ref{eq.3.34}).

Carrying out the integrals in Eq.~(\ref{eq.3.34}) and putting
$\bm{p'} = \bm{p}$ once again, we end up with
\begin{equation}
\la\psi _{p,\bm{s}}|
O(\bm{x},0)|\psi_{p,\bm{s}}\ra=\left[f_{1}(p,s) + \frac{i
\bm{x}}{2 \lambda ^2}\cdot ({\bm{ f}_{2}(p,\bm{s}) - \bm{f}_{3}
(p,\bm{s}))}\right] \frac{\bar{C}}{\lambda ^{3}} e ^{-x^2/2\lambda
^2} \label{eq.3.37}
\end{equation}
where, via Eq.~(\ref{eq.3.31})
\begin{equation}
\frac{\bar{C}}{\lambda ^{3}}= \frac{N^{2}}{(\sqrt2 \lambda)^6 }=
\frac{1}{(2 \pi)^{3/2} }\frac{1}{2 p_0} \frac{1}{\lambda^3}
\label{eq.3.38}
\end{equation}
so that \( \bar{C}= \frac{1}{(2\pi)^{3/2}} \frac{1}{2 p_0}\).

We now apply this to the case where $O(x)$ is the canonical
energy-momentum density $T_C^{\mu\nu}(x)$.  The most general
structure of the matrix elements of the conserved operator
$T_C^{\mu\nu}(0) $ is
\begin{eqnarray}
\la q',\bm{s}|T^{\mu \nu}_C(0)|q,\bm{s}\ra & = & \bar{u}(q',
\bm{s}) \{ \mathbb{G}(q^{\mu}q^{\nu} + q'^{\mu}q'^{\nu}) \nonumber \\
& & + \mathbb{H}(q^{\mu}q'^{\nu} + q^{\nu}q'^{\mu}) + M\mathbb{S}
[(q + q')^{\mu}
\gamma^{\nu} + (q+q')^{\nu} \gamma^{\mu}] \nonumber \\
& & + (q \cdot q'-M^{2})(\mathbb{G}-\mathbb{H})g^{\mu \nu} \nonumber \\
& & + M\mathbb{A}[(q + q')^{\mu} \gamma ^{\nu} -(q + q')^{\nu}
\gamma ^{\nu}] \} u(q, \bm{s}) \label{eq.3.39}
\end{eqnarray}
where the $u(q, \bm{s}), u (q', \bm{s})$ are the usual canonical
Dirac spinors normalized to $\bar{u} u = 1$ and $\mathbb{G},
\mathbb{H}, \mathbb{S}$ and $\mathbb{A}$ are Lorentz scalars. Note
that all terms, except the $\mathbb{A}$-term, are symmetric in
$\mu \leftrightarrow \nu$.  In order to identify the function
$f_1$ and the vector function $\bm{f_{2} - f_{3}}$ in
Eq.~(\ref{eq.3.33}) we have to expand Eq.~(\ref{eq.3.39}) about
$\gb{q = p}$ and $\gb{q' = p}$. The Gordon decomposition
\begin{equation}
\bar{u}(q') \gamma ^{\mu} u (q) =  \frac{(q +q')^{\mu}}{2M}
\bar{u}(q') u (q)+ \frac{i(q'-q)_\rho}{2 M} \bar {u}(q')
\sigma^{\rho \mu} u(q) \label{eq.3.40}
\end{equation}
is helpful, because to the accuracy required we can replace $q$
and $q'$ by $p$ in the factors multiplying $(q'-q)_\rho$ and then
use
\begin{equation}
\bar{u}(p, \bm{s}) \sigma^{\rho\mu} u (p, \bm{s}) = \frac{1}{M}
\epsilon ^{\rho \mu \alpha \beta}\mathcal{S}_{\alpha} p_{\beta}
\label{eq.3.41}
\end{equation}

where $\mathcal{S}$ is given by Eq.~(\ref{covcompS}) and the
convention is $\epsilon ^{0123} = + 1$.  Then
\begin{eqnarray}
\la q',\gb{s}|T_C^{\mu\nu}(0)|q,\gb{s}\ra & = &
[\mathbb{B}(q^{\mu}q^{\nu}+q'^{\mu}q'^{\nu}) + (q \cdot q'
-M^{2})(\mathbb{B}-\mathbb{C})
g^{\mu\nu} \nonumber \\
  & &+ \mathbb{C} (q^{\mu}q'^{ \nu} + q^{\nu} q'^{\mu})] \bar{u}(q', \gb{s})
u(q, \gb{s}) \nonumber  \\
& &+ \frac{i(q'-q)_\rho}{M}( \mathbb{S}[p^{\mu} \epsilon ^{\rho
\nu \alpha
\beta}+ p^{\nu} \epsilon ^{\rho \mu \alpha \beta}] \nonumber  \\
& &+ \mathbb{A} [p^{\mu} \epsilon^{\rho \nu \alpha \beta} -
p^{\nu}\epsilon^{\rho \mu \alpha \beta} ])\mathcal{S}_{\alpha}
p_{\beta} \label{eq.3.42}
\end{eqnarray}
where $\mathbb{B} = \mathbb{G} + \mathbb{S}$ and $\mathbb{C} =
\mathbb{H} + \mathbb{S}$.

Putting
\begin{eqnarray}
u (q, \gb{s}) & = & u (p, \gb{s}) +(q-p)^k u^k (p,
\gb{s})\nonumber\\
\bar{u}(q', \gb{s})& = & \bar{u} (p, \gb{s}) + (q'-p)^k \bar{u}^k
(p, \gb{s}) \label{eq.3.43}
\end{eqnarray}
where $u^k (p, \gb{s}) = \partial/\partial p^k u (p, \gb{s})$ and
similarly for $\bar{u}^k$, we find, after much algebra, that for
$T^{\mu\nu}_{C} (0)$  the function $f_{1}$ and the $k^{th}$
component of $\bm{f_{2} -f_{3}}$, needed in Eq.~(\ref{eq.3.37})
are:
\begin{equation}
f_{1} = 2 \mathbb{D} p^{\mu} p^{\nu} \label{eq.3.44}
\end{equation}
where
\begin{equation}
\mathbb{D} = \mathbb{B} + \mathbb{C} \label{eq.3.45}
\end{equation}
and
\begin{eqnarray}
(f_{2} - f_{3})^{k} & = & 2 \mathbb{D}\, p^{\mu} p^{\nu} [\bar {u}
(p, \bm{s}) u^{k} (p, \bm{s}) -
\bar{u} ^{k} (p, \bm{s}) u (p, \bm{s})] \nonumber\\
& & -2 i \frac{(\mathbb{S} + \mathbb{A})}{M} p^{\mu}
\left[\frac{p^{k}}{p_{0}} \epsilon ^{0 \nu \alpha \beta} -
\epsilon ^{k \nu \alpha \beta}\right] \mathcal{S}_{\alpha}
p_{\beta} \nonumber \\
& & -2 i \frac{(\mathbb{S} -\mathbb{A})}{M} p^{\nu}
\left[\frac{p^{k}}{p_0} \epsilon^{0 \mu \alpha \beta} - \epsilon
^{k \mu \alpha \beta}\right] \mathcal{S}_{\alpha} p_{\beta}
\label{eq.3.46}
\end{eqnarray}
so that from Eq.~(\ref{eq.3.37})
\begin{eqnarray}
\la\psi_{p, \bm{s}} | T_{C}^{0 j} (\bm{x}, 0) | \psi_{p,
\bm{s}}\ra & =& \frac{\bar{C}}{\lambda^3}
e^{- \bm{x} ^2/2 \lambda ^2}\bigg\{ 2 \mathbb{D} p^{0} p^{j}  \nonumber \nonumber \\
&& \left. + i \frac{x^{k}}{2 \lambda^{2}} \bigg[2 p^{0}p^{j}
(\bar{u}(p, \bm{s}) u^{k}(p, \bm{s})
-\bar{u}^{k}(p, \bm{s}) u(p, \bm{s}))   \right. \nonumber \\
&& \left. \left. -2 i \frac{(\mathbb{S}+ \mathbb{A})}{M}
p^{0}(\frac{p^{k}}{p^0} \epsilon ^{0 j \alpha
\beta}-\epsilon^{kj\alpha \beta}) \mathcal{S}_{\alpha} p_{\beta }
+ 2 i \frac{(\mathbb{S}-\mathbb{A})}{M} p^{j} \epsilon^{k 0 \alpha
\beta} \mathcal{S}_{\alpha} p_{\beta}\right] \right\}
\nonumber \\
\label{eq.3.47}
\end{eqnarray}

Consider now the matrix element \footnotemark[2]
\footnotetext[2]{Note that here, in the general field theoretic
case, $M_{\rm orb}$ is actually the sum of the quark orbital
angular momentum plus the full angular momentum of the gluons}
\begin{eqnarray}
\la\psi_{p, \bm{s}}|M^{i j}_{\rm orb}|\psi_{p,\bm{s}}\ra & = &
\la\psi_{p, \bm{s}}| \int d^{3}x
M^{0ij}_{\rm orb}(\bm{x}, 0)|\psi_{p, \bm{s}}\ra \nonumber \\
& = & \int d^{3}x x^{i}\la\psi_{p, \bm{s}}|T_C^{0 j}(\bm{x},
0)|\psi_{p, \bm{s}}\ra - (i \leftrightarrow j) \label{eq.3.48}
\end{eqnarray}
We see that integrating over space kills the first term in
Eq.~(\ref{eq.3.47}), and the second only contributes if $k=i$.
The spatial integral is then simply
\begin{equation}
\frac{\bar{C}}{\lambda^{3}}\int d^{3} x (x^{i})^{2} e^{-x^2/2
\lambda^ 2} = \frac{\lambda^{2}}{2 p_0} \label{eq.3.49}
\end{equation}
where we have used Eq.~(\ref{eq.3.38}) for $\bar{C}$.  The
$\lambda ^{2}$ in Eq.~(\ref{eq.3.49}) cancels the remaining
$1/\lambda^2$ in Eq.~(3.42), so that we can take the limit
$\lambda \rightarrow \infty$ and have
\begin{eqnarray}
\la\psi_{p, \bm{s}}|M^{i j}_{\rm orb}|\psi_{p,\bm{s}}\ra & = &
\frac{i}{2 p_{0}}\{\mathbb{D}
p^{0}p^{j} [\bar{u}(p, \bm{s}) u^{i}(p, \bm{s}) \nonumber\\
& &-\bar{u}^{i}(p, \bm{s}) u(p, \bm{s})] -\frac{i \mathbb{S}}{M}
[p^{i} \epsilon^{0 j \alpha \beta}- p_{0} \epsilon^{i j
\alpha \beta} + p^{j}\epsilon^{0i\alpha \beta}] \mathcal{S}_{\alpha} p_{\beta} \nonumber  \\
& &-\frac{i \mathbb{A}}{M} [p^{i} \epsilon^{0 j \alpha
\beta}-p_{0}\epsilon^{i j \alpha \beta} - p^{j}\epsilon^{0i\alpha
\beta }] \mathcal{S}_{\alpha} p_{\beta}\} -(i \rightarrow j)
\label{eq.3.50}
\end{eqnarray}
Part of the term multiplying $\mathbb{S}$ is symmetric under $(i
\leftrightarrow j)$, so cancels out.  The other terms multiplying
$\mathbb{S}$ and $\mathbb{A}$ are antisymmetric so just get
doubled under $(i \leftrightarrow j)$.  Also
\begin{eqnarray}
\bar{u}(p, \bm{s})u^{i}(p, \bm{s}) - \bar{u}^{i}(p, \bm{s})
u (p, \bm{s}) \nonumber \\
& = & \bar{u}(p,\bm{s})u^{i} (p, \bm{s}) -[\bar{u} (p,
\bm{s})u^{i} (p, \bm{s})]^* \nonumber \\
& = &  2 i \, Im [\bar{u}(p, \bm{s}) u^{i} (p, \bm{s})]
\label{eq.3.51}
\end{eqnarray}
Thus eq.~(\ref{eq.3.50}) becomes
\begin{eqnarray}
\la\psi_{p, \bm{s}}|M^{i j}_{\rm orb}|\psi_{p, \bm{s}}\ra & = & -
\mathbb{D}[p^{j} Im [\bar{u}(p, \bm{s})
u^{i} (p, \bm{s})] -(i \leftrightarrow j)] \nonumber \\
& & -\frac{\mathbb{S}}{M} \epsilon^{i j \alpha \beta}
\mathcal{S}_{\alpha} p_{\beta} + \frac{\mathbb{A}}{M p_{0}} (p^{i}
\epsilon^{0 j \alpha \beta} - p^{j} \epsilon^{0 i \alpha \beta }-
p_{0} \epsilon^{i j \lambda \beta}) \mathcal{S}_{\alpha} p_{\beta}
\label{eq.3.52}
\end{eqnarray}
Finally we expand the spinors as in eq.~(\ref{eq.3.43}) and find
that
\begin{equation}
Im [\bar{u}(p, \bm{s})u^{i}(p, s)] = \frac{1}{2M (p_{0} + M)}
\epsilon^{i l m} p_{l} s_{m} \label{eq.3.53}
\end{equation}
and comparing Eq.~(\ref{eq.27}) with $A_0(0) = 2$ [(see
Eq.~(\ref{eq.30})] with Eq.~(\ref{eq.3.42}), we see that
$\mathbb{B} + \mathbb{C} = 1$ and so, via Eq.~(\ref{eq.3.45}),
$\mathbb{D} = 1$.  Thus
\begin{eqnarray}
\la\psi_{p, \bm{s}}|M^{i j}_{\rm orb}|\psi_{p, \bm{s}}\ra & = &
\frac{1}{2M (p_{0} +
M)}[p^{i}(\bm{p} \times \bm{s})^{j}- p^{j}(\bm{p} \times \bm{s})^{i}] \nonumber\\
& & - \frac{\mathbb{S}}{M} \epsilon^{i j \alpha \beta}
\mathcal{S}_{\alpha} p_{\beta} + \frac{\mathbb{A}}{Mp_{0}} (p^{i}
\epsilon^{0 j \alpha \beta} - p^{j} \epsilon^{0 i \alpha \beta} -
p_{0}\epsilon^{ij\alpha \beta} ) \mathcal{S}_{\alpha} p_{\beta}
\label{eq.3.54}
\end{eqnarray}

We shall return to consider the interpretation of this result
presently.  First, though, we use it to deduce the structure of
the matrix elements of $J^{i j}$, which, by Eq.~(\ref{eq.2.12}),
are the same as those of $M^{i j}$.  The latter are built from the
symmetric energy momentum density $T^{\mu\nu} = \half (T^{\mu
\nu}_{C} + T^{\nu \mu}_{C})$.  Since in Eq.~(\ref{eq.3.39}) all
terms are symmetric in $\mu, \nu$  except the $\mathbb{A}$-term
which is antisymmetric, we obtain the matrix elements of
$T^{\mu\nu}$ by simply putting $\mathbb{A} = 0$ in
Eq.~(\ref{eq.3.54}). Thus
\begin{eqnarray}
\la\psi_{p, \bm{s}}|M^{i j}|\psi_{p, \bm{s}}\ra & = & \frac{1}{2M(
{p_{0}} + M)} [p^{i} (\bm{p} \times \bm{s})^{j} - p^{j} (\bm{p}
\times \bm{s})^{i}] - \frac{\mathbb{S}}{M} \epsilon^{i j \alpha
\beta} \mathcal{S}_{\alpha} p_{\beta}. \label{eq.3.55}
\end{eqnarray}

We obtain the value of $\mathbb{S}$ by choosing a wave-packet with
$\bm{p} = (0, 0, p)$ and $\bm{s} = (0, 0,1)$.  This is then a
helicity state $|\psi_{1/2}\ra$    and should be an eigenstate of
$J_{z}$ with eigenvalue $1/2$.  Hence for this state
\begin{equation}
\la\psi_{1/2}|J_{z}|\psi_{1/2}\ra =
\la\psi_{1/2}|M^{12}|\psi_{1/2}\ra = 1/2 \label{eq.3.56}
\end{equation}
which implies
\begin{equation}
\frac{1}{2} = - \frac{\mathbb{S}}{M} \epsilon^{1203}
(\mathcal{S}_{0} p_{3} - \mathcal{S}_{3} p_{0}) =
\frac{\mathbb{S}}{M} (\mathcal{S}_{0} p^{3} - \mathcal{S}^{3}
p_{0}) \label{eq.3.57}
\end{equation}
Now from Eq.~(\ref{covcompS}), for the present case,
\begin{eqnarray}
  \mathcal{S}_{0} & = & p/M \nonumber\\
  \mathcal{S}^{3} & = & 1 + \frac{p^{2}}{M(p_{0} + M)} = p_{0}/M
\label{eq.3.58}
\end{eqnarray}
Hence Eq.~(\ref{eq.3.57}) becomes,
\begin{equation}
\frac{1}{2} = \frac{\mathbb{S}}{M^{2}} (p^{2}- p^{2}_{0}) =
-\mathbb{S} \label{eq.3.59}
\end{equation}
Putting this into Eq.~(\ref{eq.3.55}) gives
\begin{eqnarray}
\la\psi_{p, \bm{s}}|M^{i j}|\psi_{p, \bm{s}}\ra & = &
\frac{1}{2M}\left\{\frac{1}{p_{0} + M}
[p^{i}(\gb{p \times s})^{j} - p^{j} (\gb{p \times s})^{i}] \right.\nonumber\\
&&  +  \epsilon^{i j \alpha \beta}  \mathcal{S}_{\alpha}
p_{\beta}\bigg\} \label{eq.3.60}
\end{eqnarray}

Let us first compare this result to what we obtained in
Eq.~(\ref{eq.71}), Eq.~(\ref{eq.75}) for the relativistic
quantum-mechanical Dirac particle.  The last term in
eq.~(\ref{eq.3.60}) can be written
\begin{eqnarray}
  \epsilon^{i j \alpha \beta}  \mathcal{S}_{\alpha} p_{\beta} & = &
\epsilon^{i j 0\beta}[ \mathcal{S}_{0} p_{\beta}-
\mathcal{S}_{\beta}
p_{0}] \nonumber  \\
& = & \epsilon^{i j k} \left[p_{0} \mathcal{S}^k - \frac{\gb{p
\cdot
s}}{M} p^{k}\right] \nonumber  \\
& = & \epsilon^{i j k} \left[p_{0} (s^{k} + \frac{\gb{p \cdot
s}}{M(p_{0}+M)} p^{k}) -
\frac{\gb{p \cdot s}}{M}\,p^{k}\right] \nonumber  \\
& = & \epsilon^{i j k}\left[ p_{0} s^{k} - \frac{\gb{p \cdot
s}}{p_{0} + M} \,p^{k}\right] \label{eq.3.61}
\end{eqnarray}
Putting this into Eq.~(\ref{eq.3.60}) we have, finally,
\begin{eqnarray}
\la\psi_{p, \bm{s}}|M^{i j}|\psi_{p, \bm{s}} \ra & = &
\frac{\epsilon^{i j k}}{2 M} \left\{\frac{1}{p_{0} + M} [\gb{p
\times (p \times s)}]^{k} + p_{0}s^{k} -
\frac{\gb{p \cdot s}}{p_{0} + M} p^{k}\right\} \nonumber \\
& = & \frac{\epsilon^{i j k}}{2 M} \left(p_{0}-
\frac{\bm{p}^{2}}{p_{0} + M}\right)
s^{k} \nonumber \\
& = & \frac{\epsilon^{i j k}}{2} s^{k} \label{eq.3.62}
\end{eqnarray}
This is exactly the result found in Eq.~(\ref{eq.71}),
Eq.~(\ref{eq.75}). Taken in conjunction with Eqs.~(\ref{eq.09}),
(\ref{eq.10}) and (\ref{eq.2.12}), the result Eq.~(\ref{eq.3.62})
corresponds to the first term in Eq.~(\ref{eq.5.011}). The
derivative term in the latter equation is missing here because of
our use of a wave packet. We shall compare Eq.~(\ref{eq.3.62}) to
the results of [3] and [4] in Section~6.

Let us return now to the canonical form of the angular momentum.
It is shown in [4] for QCD that aside from the terms which vanish
by virtue of the equations of motion or which give no contribution
to forward matrix elements
\begin{equation}
M^{\mu\nu\lambda}{(x)} = M^{\mu\nu\lambda}_{\rm orb}(x) +
M^{\mu\nu\lambda}_{\rm axial}(x) \label{eq.3.63}
\end{equation}
where in the convention $\epsilon^{0123} = + 1$, the axial density
is
\begin{equation}
M^{\mu\nu\lambda}_{\rm axial}(x) = \half
\epsilon^{\mu\nu\lambda\alpha} \bar{\psi}(x)\gamma_{5}
\gamma_{\alpha} \psi(x) \label{eq.3.64}
\end{equation}
and a sum over the color labels of the quark fields is implied.

 Note that
Eq.~(\ref{eq.3.64}) is not the same as the spin density in
Eq.~(\ref{eq.60}), for arbitrary values of $\mu,\nu,\lambda$.  But
Eq.~(\ref{eq.60}) and Eq.~(\ref{eq.3.64}) are equal if $\mu \ne
\nu , \lambda$.  Thus
\begin{equation}
M^{0ij}_{\rm axial}(x) = M^{0ij}_{\rm{spin} \textit{f}}\,(x)
\label{eq.3.65}
\end{equation}
 where ``spin $f$" means the fermionic spin terms in the angular momentum operator.
This is the basis for the result in Eq.~(\ref{eq.2.12}).  The
axial density is a local operator, so that
\begin{equation}
\la p', \bm{s}|\int d^{3} x M^{\mu\nu\lambda}_{\rm axial}(x)|p,
\bm{s}\ra = (2\pi)^{3} \delta^{3}(\bm{p}' - \bm{p}) \la p',
\bm{s}|M^{\mu\nu\lambda}_{\rm axial} (0)|p, \bm{s}\ra
\label{equation}
\end{equation}
The only possible structure for the matrix element on the RHS is
\begin{equation} \la p, \bm{s} |M^{\mu\nu\lambda}_{\rm axial}(0)|p, \bm{s}\ra
= - a_{0} \epsilon^{\mu\nu\lambda\rho} \mathcal{S}_{\rho}
\label{eq.3.67}
\end{equation}
where $a_{0}$ is known as the axial charge. Thus for the
expectation value
\begin{equation}
\frac{\la p', \bm{s}|\int d^{3} x M^{\mu\nu\lambda}_{\rm axial}
(x, 0)|p, \bm{s}\ra}{\la p',\bm{s}|p, \bm{s}\ra} = -
\frac{a_{0}}{2 p_{0}}
\epsilon^{\mu\nu\lambda\rho}\mathcal{S}_{\rho} \label{eq.3.68}
\end{equation}
 From Eq.~(\ref{eq.3.65}) we thus have for the spin density
\begin{equation}
\frac{\la p', \bm{s}|\int d^{3}x M^{0ij}_{\rm{spin}
\textit{f}}\,(x, 0)|p, \bm{s}\ra}{\la p', \bm{s}|p, \bm{s}\ra} = -
\frac{a_{0}}{2 p_{0}} \epsilon^{i j k} \mathcal{S}_{k}
\label{eq.3.69}
\end{equation}
After a little algebra, using Eq.~(\ref{covcompS}), we write the
fermionic part of the spin vector $\textbf{S}^f $ of the nucleon
in terms of the independent vectors and $\bm{p}$ and
$\bm{s}$:\footnotemark[2] \footnotetext[2]{Note that we are
deliberately not calling $\textbf{S}^f $ the quark spin
contribution to the nucleon spin. In an interacting theory with an
anomaly the connection between $\textbf{S}^f $ and the spin
carried by the quarks is subtle and will be discussed in
Section~7.}
\begin{equation}
S^f_k = \frac{a_{0}}{2p_{0}} \left[M s_k + \frac{(\bm{p} \cdot
\bm{s})}{p_{0} + M} p_k \right] \label{eq.3.70}
\end{equation}
Similarly, for the vector representing the fermionic orbital
angular momentum vector plus the full gluon angular momentum,
Eq.~(\ref{eq.3.54}) yields
\begin{equation}
L^f_k + J^G_k = \left(\frac{1}{2} - \mathbb{A}
\frac{M}{p_{0}}\right) s_k - \mathbb{A} \frac{\gb{p \cdot
s}}{p_{0}(p_{0} + M)} p_k \label{eq.3.71}
\end{equation}
Adding Eq.~(\ref{eq.3.70}) and eq.~(\ref{eq.3.71}) gives
\begin{equation}
\textbf{J} = \left[\frac{1}{2} - \frac{M}{p_{0}} (\mathbb{A} -
\frac{a_{0}}{2})\right] \gb{s} - (\mathbb{A} -
\frac{a_{0}}{2})\frac{\gb{p \cdot s}}{p_{0}(p_{0}+M)} \gb{p}
\label{eq.3.72}
\end{equation}
Then Eq.~(\ref{eq.3.62}) implies that
\begin{equation}
\mathbb{A}= \frac{a_{0}}{2} \label{eq.3.72}
\end{equation}
and the contribution from the axial charge cancels against the
antisymmetric term in the fermionic orbital plus full gluonic
angular momentum.

This supports the claim  in [4] that the axial contribution
totally cancels out when taking forward matrix elements of
Eq.~(\ref{eq.3.63}), at least for the $0ij$ elements of $M^{\mu
\nu \lambda}$. So although the proof given in [4] is invalid
because it suffers from the problems mentioned in Comments~1 and
3, it is correct in  the contention that the axial anomaly cancels
out of the angular momentum sum rules.\footnotemark[2]
\footnotetext[2]{This does not mean that the spin carried by the
quarks does not contribute to the spin of the nucleon. As
explained in \cite{Leader} $a_0$ contains also an anomalous gluon
contribution.}As the authors of \cite{ref.04} realize (cf. their
Eq.~(21)), this must occur: because of the non-renormalization of
the energy-momentum tensor,  the total angular momentum must not
be anomalous. See also \cite{Hoodbhoy}.

  \section{AN INDEPENDENT APPROACH}

We have argued that the forms for the angular momentum tensor
given in [3] and [4] are incorrect, in spite of their appearance
of explicit Lorentz invariance. On the other hand our procedure
has led to the surprising result Eq.~(\ref{eq.3.62}) that for a
spin $\frac{1}{2}$ particle described by a canonical spin state
 \beq
 \la{p', \bm{s}}| M^{ij}|{p,
\bm{s}}\ra = 2\, p_0 \left(\epsilon^{ijk}\bm{s}_k/2 + i p^i
\partial^j -i p^j \partial^i \right) \, (2
\pi)^3\delta^{(3)}(\bm{p'}-\bm{p}), \label{B} \eeq though the
derivatives of the delta-functions shown here disappeared in our
wave-function treatment, via partial integration, for the kind of
wave packet chosen in Section~5.
  The key part of Eq.~(\ref{B}), namely the first term on the RHS,
   was derived in several different ways, where always careful
attention was paid to the definition of matrix elements required
by the compound nature of the angular momentum density operator.
However, it does not share the explicit Lorentz invariance of the
form Eq.~(\ref{eq.37}) and so one may question its correctness.

In this Section we will confront this issue  by deriving
Eq.~(\ref{B}) by a totally different method, \textit{valid for
arbitrary spin}, which is based on the rotational properties of
states, and which circumvents completely the use of the
energy-momentum tensor, and demonstrate that Eq.~(\ref{B}) is
exactly what is expected on very general grounds. We shall then
show that our result does have the correct Lorentz transformation
properties. Finally we shall derive the analogue of Eq.~(\ref{B})
for particles described by Jacob-Wick helicity states.

\subsection{Canonical spin state matrix elements}
\label{sec.4.1}

In order to utilize the rotational properties of the canonical or
boost spin states we need to display explicitly the Wigner boost
operators used in defining the states of a moving particle in
terms of the rest frame spin states quantized in the $z$-direction
\( |{0, m}\ra. \) Then the transformation properties of the states
becomes explicit.

%The state for a spin-$s$ particle at rest in an eigenstate of
%spin, with spin pointing along the $\bm{s}$ direction, are given
%by \beq { |0,\bm{s} } \ra = \mathcal{D}^{\,s}_{m s} (R(\bm{s}))
%|{0,m}\ra \label{s(m)}
 %\eeq
 %where $R(\bm{s})$ rotates a unit vector in the
%$z$-direction into $\bm{s}$ by first a rotation about $y$ and then
%a rotation about $z$, a common convention. Note that we have
%constructed an eigenstate with spin eigenvector along $\gb{s}$ and
%the state is thus completely specified even if $s\geqq 1$.

  We  will use the definitions of the Lorentz group generators given
in Weinberg \cite{Weinberg}.  In this section we will write the
3-vector operators in terms of the integrated tensor $M^{ij}$
  \begin{eqnarray}
  J_i &=& \half \epsilon_{ijk} M^{jk}  \\
  K_i &=& M^{i0}.
  \end{eqnarray}
  where on the left hand side $ i = x,y,z.$
  For a particle of mass $M$ in motion the boost (or canonical) state is defined by
  \begin{equation}
|{p, m}\ra =B(\bm{v})|0,m \ra =\exp(i \zeta \, \hat{\bm{p}} \cdot
\bm{K}) |0, m \ra,
 \label{booststate}
\eeq where $\bm{v} = \bm{p}/p_0,  \, \cosh{\zeta} =  p_0/M$, and
$\hat{\bm{p}}$ is the unit vector along $\bm{p}$.

  Now consider a rotation about axis-$i$ through an angle $\beta$. The
unitary operator which effects this is given in terms of the
angular momentum operator $J_i$:
  \beq
R_i(\beta)= \exp(-i \beta J_i)
  \eeq
  and for a particle of arbitrary spin-$s$
  \beq
R_i(\beta))| {p,m}\ra = |{R_i(\beta) p, n}\ra \mathcal{D}^{\,
s}_{nm} (R_W(p,\beta)).
  \eeq
  where $R_W(p,\beta)$ is the Wigner rotation.
  In a slight abuse of notation, we will use the symbol $R$ to denote
both the unitary operator for rotations in Hilbert space and the
corresponding rotation matrix in Minkowski space. The same will be
done for boosts, $B$. For a pure rotation the Wigner rotation
$R_{W}$ is very simple
\[ R_{W}(p,\beta) =R_i (\beta) ; \]
independent of $p$ ; cf. for example \cite{Leader}. Therefore,
  \beqn
   \la{p',m'}| R_i(\beta) |{p,m}\ra & = & \la{p',m'}|R_i(\beta){ p,
n}\ra \mathcal{D}^{\, s}_{nm }(R_i(\beta)) \\
& = & 2 p_0 (2 \pi)^3 \delta^{(3)} (\bm{p}' - R_i(\beta) \bm{p})
\mathcal{D}^{\, s}_{m'm }(R_i(\beta)), \nonumber
   \eeqn
   using the conventional normalization
\beq
 \la{p',m'} |{p,m}\ra  = 2 p_0(2 \pi)^3 \delta^{(3)}(\bm{p}'-\bm{p}) \delta_{m'm}.
 \eeq
   Thus
   \beqn
   \la p',m'|J_i|p,m \ra &=&i\, \frac{\partial}{\partial \beta }  \la{p',m'}|
(R_i(\beta) |{p,m}\ra  |_{\beta = 0}\\
   &=&2p_0 (2\pi)^3 \left[ i\epsilon_{ijk} p_j
\frac{\partial}{\partial p_k}\delta_{m'm} +
i\frac{\partial}{\partial \beta }\mathcal{D}^{\, s}_{m'm
}(R_i(\beta))\big|_{\beta = 0} \right]\delta^{(3)}(\gb{p'-p}).
\label{J}
   \eeqn
   Now \cite{Rose}
   \beqn \label{C}
i\frac{\partial}{\partial \beta }\mathcal{D}^{\, s}_{m'm
}(R_i(\beta))\big|_{\beta = 0} = (\textrm{S}_i)_{m'm} \eeqn
 where the three $(2s+1)$ dimensional matrices $\textrm{S}_i$
 are the spin matrices for spin-$s$ which satisfy
\beqn \label{D}
 [ \textrm{S}_j \, , \textrm{S}_k ] = i\epsilon_{jkl} \,
 \textrm{S}_l
 \eeqn
 Thus, our final result for the matrix elements of the angular momentum
 operators for arbitrary spin, from  Eq.~(\ref{J}), becomes
 \beqn \label{JK}
\la p',m'|J_i|p,m \ra =2p_0 (2\pi)^3 \left[ \textrm{S}_i +
i\epsilon_{ijk} p_j \frac{\partial}{\partial p_k}
\right]_{m'm}\delta^{(3)}(\gb{p'-p}). \eeqn
 For spin $\frac{1}{2}$, of course, the $\textrm{S}_i$ are just $ \frac{1}{2}$
 times the Pauli matrices $ \sigma _i$. For arbitrary spin they
 are still very simple:
 \begin{eqnarray}
 (\textrm{S}_z )_{m'm} &=& m \, \delta_{m'm} \nonumber \\
 (\textrm{S}_x)_{m'm} & = & \frac{1}{2}\, [C(s,m)\, \delta_{m',m+1} + C(s,-m)\,
 \delta_{m',m-1}] \nonumber \\
 ( \textrm{S}_y)_{m'm} & = & \frac{-i}{2}\,
 [C(s,m)\,\delta_{m',m+1}- C(s,-m)\,\delta_{m',m-1}] \label{E}
 \end{eqnarray}
where
 \beqn \label{F}
 C(s,m) = \sqrt{(s-m)(s+m+1)}
 \eeqn
 For the case of spin $ \frac{1}{2}$ Eq.~(\ref{JK})
 is exactly equivalent to the result quoted in Eq.~(\ref{J11})
 and which we obtained in Section~5
 after much labor using the wave packet approach.
 It is completely general. The second term will
vanish if integrated over symmetric wave packets. However it must
be kept for analyzing the transformation properties, as we will
see, and must, as usual, always be interpreted in the sense of
partial integration. It is very easy to verify that the form
Eq.(\ref{JK}) satisfies the usual commutation relation relations
and so is consistent with rotational invariance.

Combining the result Eq.(\ref{JK}) for the case of spin $
\frac{1}{2}$ with Eq.~(\ref{5.022}) leads directly to the result
quoted in Eq.~(\ref{eq.5.011}).

 \subsection{Lorentz invariance}
 We shall now demonstrate that, despite appearances to the
 contrary, Eq~(\ref{JK}) is consistent with Lorentz invariance.
 Because of the complicated algebra involved, we shall present
 the proof just for spin $ \frac{1}{2}$ of mass $M$.
 Under Lorentz transformations,
$M^{i0}=-M^{0i} = K_i$ are brought in so we need the matrix
elements of $K_i$. These can be obtained just as those for the
angular momentum. For a boost of magnitude and direction
$\bm{\omega}$
   \beq
  B(\bm{\omega})= \exp(i \bm{\omega} \cdot \bm{K}).
  \eeq
 Here we are using a natural shorthand for a boost Eq.~(\ref{booststate})
  with velocity $\bm{v} = \hat{\bm{\omega}} \tanh(\omega) \approx  \bm{\omega}$
   for small $|\bm{\omega}|$.
   Proceeding as before we have
    \beq
   \la{p',m'}| B(\bm{\omega}) |{p,m}\ra = \la{p',m'}|{B(\bm{\omega})p, n}\ra
\mathcal{D}^{1/2}_{nm }(R_W(\bm{p},\bm{\omega})).  \label{boost}
   \eeq

   The Wigner rotation for this case is defined by
   \beq
   B(B(\bm{\omega}) p) R_W(\bm{p}, \bm{\omega}) = B(\bm{\omega})B(\bm{p}) \label{Wigner}
   \eeq
   and is much more complicated than for rotations. However, we will only
   need the Wigner angle for small Lorentz
transformations for our discussion. By doing the matrix
multiplication explicitly for small $\bm{\omega}$ we find the
Wigner rotation is given, in magnitude and direction, by
   \beq
  \gb{ \alpha}=\frac{\bm{p} \times \bm{\omega}}{p_0 +M}.\label{alpha}
   \eeq
   In this case the Wigner rotation angle depends on $\bm{p}$ and this
leads to the necessarily more complicated  commutation properties
involving the boost generators.
   Differentiating  Eq.(\ref{boost}) with respect to $\omega$ and then
putting $\omega=0$ produces
  \beq
  \la{p',m'}|K_i|{p,m}\ra =2p_0 (2\pi)^3 \left[-i p_0 \partial_i +
\frac{1}{2} \frac{(\bm{p} \times
\bm{\sigma})_i}{p_0+M}\right]_{m'm}\delta^{(3)}(\bm{p}'-\bm{p}).
\label{K}
  \eeq

  We will now verify that the matrix elements of $J_i$ and $K_i$
transform among each other correctly. The Lorentz transformation
of $J_i$ under a boost $\bm{\omega}$ requires that
\begin{eqnarray}
\la{p',m'}|J_i|{p,m}\ra & =
&\la{B(\bm{\omega})p',n'}|B(\bm{\omega}) J_i  \,
B(\bm{\omega}) ^{-1}|{B(\bm{\omega})p,n}\ra   \nonumber \\
& &  \mathcal{D}^{1/2*}_{n',m'}(R_W(\bm{p} ', \bm{\omega}))
\mathcal{D}^{1/2}_{n,m}(R_W(\bm{p} , \bm{\omega})). \label{boost
of J}
\end{eqnarray}
Notice that the two Wigner rotations appearing in Eq.(\ref{boost
of J}) are not the same as long as $p \neq p'$. It is essential to
keep this is mind because of the derivatives of $\delta$-functions
that enter. If we write this out to terms linear in $\bm{\omega}$
and use \beq
  [J_i, K_j] = i \epsilon_{ijk} K_k
  \label{eq.4.18a}
\eeq Eq.(\ref{boost of J}) becomes
\begin{eqnarray}
\la{p',m'}|J_i|{p,m}\ra & = &\la {B(\bm{\omega})p',n'}| J_i
+\epsilon_{ijk} \omega_j K_k |{B(\bm{\omega}) p, n}\ra  \nonumber
\\ & + &\left(1 - i\half \epsilon_{abc} \sigma_a\frac{\omega_b
p'_c}{p_0+M}\right)_{m'n'}\left(1 + i\half \epsilon_{abc}
\sigma_a\frac{\omega_b p_c}{p_0+M}\right)_{nm} \label{J-trans}.
\end{eqnarray}
We must also expand the matrix element to first order in $\omega$.
Using Eq.(\ref{K}) and Eq.~(\ref{JK}) with $ \textrm{S}_i$
replaced by $ \sigma_i/2$, and recalling that $p_0
\delta^{(3)}(\bm{p}'-\bm{p}) = p'_0 \delta^{(3)}(\bm{p}'-\bm{p})$
is invariant under Lorentz transformations we find that
\begin{eqnarray}
\la {B(\bm{\omega})p',n'}| J_i + \epsilon_{ijk} \omega _j K_k
|{B(\bm{\omega}) p, n}\ra  & = & \left[ i \epsilon_{ijk}p_j
\partial_k  + \half \sigma_i + \half \frac{(\bm{\omega} \times
(\bm{p} \times \bm{\sigma}) )_i}{p_0 +M}\right]_{n'n} \nonumber \\
& &2 p_0(2 \pi)^3 \delta^{(3)}(\bm{p}'-\bm{p})
\label{matrixelement} .
   \end{eqnarray}
If one inserts Eq.(\ref{matrixelement}) into Eq.(\ref{J-trans})
and evaluates the terms proportional to $\omega$ one easily finds
three components: one coming from the second term in
Eq.(\ref{matrixelement})  combined with the rotation matrices in
Eq.(\ref{J-trans}), one coming from the third term in
Eq.(\ref{matrixelement}) , and one from the first term in
Eq.(\ref{matrixelement}) acting on the rotation matrices in
Eq.(\ref{J-trans}). It is essential to carry out the derivatives
in this last component {\em before} setting $p'=p$:
\begin{eqnarray}
& & \frac{(\bm{\omega} \times (\bm{p} \times
\bm{\sigma}))_i}{2(p_0 +M)} - i \frac{[\bm{\sigma} \cdot
(\bm{\omega} \times \bm{p} ),\sigma_i]}{4(p_0 +M)}
\nonumber \\
& & + i \epsilon_{ijk} p_j \partial _k \left[1 - i
\frac{\bm{\sigma} \cdot (\bm{\omega} \times \bm{p}
')}{2(p_0+M)}\right]\left[1 + i \frac{\bm{\sigma} \cdot
(\bm{\omega} \times \bm{p})}{2(p_0+M)}\right]
   \end{eqnarray}
    Use of the commutation relations of the Pauli matrices and the
vector double cross product identity shows that the sum of these
three pieces vanishes leaving just the matrix element of $J_i$ as
was to be shown in order to satisfy Eq.(\ref{boost of J}) or
(\ref{J-trans}).

    Using the same techniques, one can show that the matrix elements of $K_{i}$  as given in
Eq.(\ref{K}) also transform correctly under boosts and so
Eq.(\ref{JK})   and Eq.(\ref{K}) form a representation of the
Lorentz group.

\subsection{Helicity state matrix elements}
\label{sec.4.2}

    We now turn to the case of helicity states which have some rather
surprising properties. One can proceed just as here; the main
difference is that the Wigner rotation becomes a Wick helicity
rotation, always about the $z$-axis. This simplifies things
somewhat; all the complication is in calculating the angle that
results, the analog of Eq.(\ref{alpha}). The result is also
convention dependent, depending on whether one uses the original
Jacob and Wick definition \cite{JacobWick} or the later one due to
Wick \cite{Wick} \, [see Eqs.~(\ref{lbdj}) and ~(\ref{przo})] . We
give here the result for the first case. The result of this messy
calculation is that, for $ \bm p = (p,\theta,\phi) $
     \beq
     \la{p',\lambda'}|J_{i}|{p,\lambda}\ra _{JW}= (2 \pi)^{3} 2 p_{0 }
\left[\lambda \eta_{i}+ i (\bm{p} \times \bm{\nabla})_{i}\right]
\delta^{(3)}(\bm{p}'-\bm{p}) \delta_{\lambda' \lambda}
\label{helicity1}
     \eeq
     where
     \beq
     \eta_{x}= \cos(\phi)\tan(\theta/2),\qquad \eta_{y}= \sin(\phi)
\tan(\theta/2), \qquad \eta_{z}=1. \label{helicity2}
     \eeq
     Although these components look a little odd---the singularity at
$\theta= \pi$ results from the ambiguity of Jacob and Wick
helicity states at that point---it is easy to verify some
important properties : they are manifestly diagonal in $\lambda$,
which is required since rotations preserve the helicity, and they
satisfy
 \beq \label{hel1}
 \la{p',\lambda'}|\hat{\bm{p}} \cdot
\bm{J}|{p,\lambda}\ra _{JW} = \lambda \; 2 p_{0 }\,(2 \pi)^{3}
\delta^{(3)}(\bm{p}'-\bm{p}) \;\delta_{\lambda' \lambda}
 \eeq
     and no orbital angular momentum piece survives as expected.

     It is enlightening to consider these amplitudes from a different
direction: comparing the definitions of canonical (boost) states
to helicity states we have for the case of spin $ \frac{1}{2}$
     \beqn
    | {p,\bm{s}}\ra &=& |{p,m}\ra  \mathcal{D}^{\, 1/2}_{m 1/2}(R(\bm{s})) \nonumber  \\
    &=& |{p,\lambda}\ra _{JW } \mathcal{D}^{\, 1/2}_{\lambda m
}(R^{-1}(\bm{p}))  \mathcal{D}^{\, 1/2}_{m 1/2 }(R(\bm{s}))  \nonumber \\
    &=& |{p,\lambda}\ra _{JW } \mathcal{D}^{\, 1/2}_{\lambda 1/2
}(R^{-1}(\bm{p}) R(\bm{s})). \label{helicitytocanonical}
    \eeqn
    This has the appearence of an ordinary unitary change of basis,
but because of the compound nature of $J_{i}$ when we apply this
to the canonical form, using the spin $ \frac{1}{2}$ version of
Eq.~(\ref{JK}), we get
    \beq
    \la{p',\lambda'}|J_{i}|{p,\lambda}\ra_{JW}=  (2 \pi)^{3}\, 2 p_{0
} \mathcal{D}^{\, 1/2}_{m \lambda }(R(\bm{p})) \mathcal{D}^{\,
1/2}_{m' \lambda '}(R(\bm{p} '))^{*} \left[i \epsilon_{ijk}
p_{j}\partial_{k} + \frac{1}{2}\, \sigma_i \right]_{m' m}
\delta^{(3)}(\bm{p}'-\bm{p}). \label{helicity3}      \eeq
      We cannot use the unitarity of the $\mathcal{D}$'s because
$\bm{p} \neq \bm{p} '$, and we must first pass the first
$\mathcal{D}^{\, 1/2}(R(\bm{p}))$ through the derivative before
setting them equal. This produces an extra term

\beq - (2 \pi)^{3} \, 2 p_{0}\, i \epsilon_{ijk}\mathcal{D}^{\,
1/2 }_{m' \lambda '}(R(\bm{p} '))^{*} p_{j}\partial_{k}
\mathcal{D}^{\, 1/2}_{m
\lambda}(R(\bm{p}))\delta^{(3)}(\bm{p}'-\bm{p}) \label{helicity4}
\eeq
 which is tedious to evaluate in the general case. The result of this
labor is identical to  Eqs.(\ref{helicity1}, \ref{helicity2}).

\section{Applications : sum rules}
\label{sect.5}

Equipped with the expressions for the matrix elements of $J_i$
derived in Sections~5 and 6 and summarised explicitly in
Eqs.~(\ref{J11}), (\ref{eq.5.041}) and (\ref{helicity21}), we will
derive the general form for angular momentum sum rules for the
nucleon and, in particular, will derive a new sum rule for
transverse polarization. This differs from the sum rule that would
be derived from Eq.~(\ref{eq.37}).
\subsection{The matrix elements of $\bm{J}$: the $J_z$ sum rule}
\label{sect.5.1}

The first term in our result Eq.~(\ref{eq.5.011}) differs from the
results of Jaffe and Manohar [3].  If we rewrite their expression
Eq.~(\ref{eq.37}) in terms of the independent vectors $\bm{p}$ and
$\bm{s}$, we find, for the expectation value
\begin{equation}
 \la J_i\ra_{JM} = \frac{1}{4Mp^0} \left\{(3p_0^{2}- M^{2}) s_i
 -\frac{3p_0 +M}{p_0+ M} (\bm{p} \cdot \bm{s}) p_i \right\}
\label{eq.5.05}
\end{equation}
to be compared to
\begin{equation}
\la J_i\ra =\half  s_i \label{eq.5.06}
\end{equation}
arising from the first term in Eq.~(\ref{eq.5.011}).  In general
these are different. However, one may easily check that if $\bm{s}
= \hat{\bm{p}}$ the Jaffe-Manohar value agrees with
Eq.(\ref{eq.5.06}), while if $\bm{s}  \perp \hat{\bm{p}}$ they are
not the same.

The agreement for  $\bm{s} = \hat{\bm{p}}$  is consistent with the
much used and intuitive sum rule
\begin{equation}
\half = \half \Delta \Sigma + \Delta G +
 \langle L^q \rangle + \langle L^G \rangle
\label{eq.5.07}
\end{equation}
based on the matrix elements of $J_z$, for a proton moving along
the $z$-axis with helicity $\lambda = \half$ (this, as explained
in Section~3, coincides with a canonical spin state), relating the
component along $\bm{p}$ of the spin and orbital angular momentum
of the quarks and gluons to the helicity of the proton.

\subsection{General structure of sum rules}
\label{sect.5.02} Consider a nucleon with momentum along $OZ$,
$\bm{p} = (0,0,p)$, in a canonical spin state with rest-frame spin
eigenvector along $\bm{s}$, where $\bm{s}$ could be longitudinal
$\bm{s}_L$ or transverse $\bm{s}_T$. Sum rules can be  constructed
by equating the expression Eq.~(\ref{J11}) for the nucleon matrix
elements $\langle \bm{p}', m' | J_i | \bm{p}, m \rangle$ with the
expression obtained when the nucleon state is expressed in terms
of the wave functions of its constituents (partons; quarks and
gluons).

There is great interest in sum rules in which the partonic
quantities can be related to other physically measurable
quantities. The classic example of this is Eq.~(\ref{eq.5.07}). We
will now investigate other similar possibilities, using
Eq.~(\ref{J11}) as the relevant starting point.

We have stressed the importance of a wave-packet approach in order
to deal with the derivative of the delta-function in the equations
above . As it happens, however, when constructing sum rules,  the
expression in terms of constituents automatically produces a term
which cancels the delta-function, irrespective of the actual model
wave-functions used.

The nucleon state is expanded as a superposition of $n$-parton
Fock states, wherein, for the purpose of showing the structure of
the sum rules, we will not display flavour and colour labels. We
use the ``Instant" form rather than the commonly used
``Light-Cone" form
 \cite{Brodsky} since it is more suitable for discussing rotational properties. We will use the original $p \rightarrow \infty$ limit in order to obtain the parton model sum rules.
\begin{eqnarray}
 | \bm{p}, m \rangle & = & [(2\pi)^3 2p_0]^{1/2} \sum_n
 \sum_{\{\sigma\}} \int \frac{d^3\bm{k}_1}{\sqrt{(2\pi)^3 2k^0_1}}
 \dots \frac{d^3\bm{k}_n}{\sqrt{(2\pi)^3 2k^0_n}}
\nonumber \\
 && \times \psi_{\bm{p}, m} (\bm{k}_1,\sigma_1, ... \bm{k}_n, \sigma_n)
 \delta^{(3)}(\bm{p} - \bm{k}_1 -  ...  - \bm{k}_n)
 |\bm{k}_1,\sigma_1,  ... \bm{k}_n, \sigma_n \rangle.
\label{eq.5.08}
\end{eqnarray}
where $\sigma_i$ denotes either the spin projection on the
$z$-axis or the helicity, as appropriate. $\psi_{\bm{p},m}$ is the
partonic wave function of the nucleon normalized so that
\begin{equation}
 \sum_{\{\sigma\}} \int d^3\bm{k}_1 \dots d^3\bm{k}_n
 | \psi_{\bm{p},m} (\bm{k}_1,\sigma_1, ... \bm{k}_n, \sigma_n)|^2
 \delta^{(3)}(\bm{p} - \bm{k}_1 -  ...  - \bm{k}_n) = {\cal P}_n.
\label{eq.5.09}
\end{equation}
with ${\cal P}_n$  denoting the probability of the $n$-parton
state.

The $n$-parton contribution is then
\begin{eqnarray}
 \langle \bm{p}', m' | J_i | \bm{p}, m \rangle ^{\rm n-parton} & = &
 (2\pi)^3 2p_0  \sum_{\sigma, \sigma'}
 \int [d^3 k'_1] \dots [d^3 k'_n][d^3 k_1] \dots [d^3 k_n]
\nonumber \\
 && \psi^*_{\bm{p}, m} (\bm{k}_1',\sigma_1', ... \bm{k}_n', \sigma_n')
 \langle \bm{k}'_1, \sigma'_1, ..., \bm{k}'_n, \sigma'_n |J_i|
  \bm{k}_1, \sigma_1, ..., \bm{k}_n, \sigma_n  \rangle
 \\
 && \psi_{\bm{p}m} (\bm{k}_1,\sigma_1, ... \bm{k}_n, \sigma_n)
 \delta^{(3)}(\bm{p}' - \bm{k}'_1- ... - \bm{k}'_n)
 \delta^{(3)}(\bm{p} - \bm{k}_1- ... - \bm{k}_n) \nonumber
\label{eq.5.10}
\end{eqnarray}
where we use the notation
\begin{equation}
 [d^3 k] = \frac{d^3 \bm{k}}{\sqrt{(2\pi)^3 2 k^0}}.
\label{eq.5.11}
\end{equation}
We take for the Fock-state matrix elements
\begin{equation}
 \langle \bm{k}'_1, \sigma'_1, ..., \bm{k}'_n, \sigma'_n |J_i|
 \bm{k}_1, \sigma_1, ..., \bm{k}_n, \sigma_n  \rangle
 = \sum_r \langle \bm{k}'_r, \sigma'_r|J_i|\bm{k}_r, \sigma_r  \rangle
\prod_{l\neq r} (2\pi)^3 2 k^0_l \delta^{(3)}(\bm{k'}_l -
\bm{k}_l)
 \delta_{\sigma'_l\sigma_l}.
\label{eq.5.12}
\end{equation}
so
\begin{eqnarray}
 \langle \bm{p}', m' | J_i | \bm{p}, m \rangle & = &
 (2\pi)^3\, 2p_0 \sum_{n, r} \sum_{\sigma_i}\sum_{\sigma'_r}
 \int [d^3 k'_r]
 \int d^3 k_1 \dots [d^3 k_r] \dots d^3 k_n \; \langle \bm{k}_r', \sigma_r'|J_i| \bm{k}_r, \sigma_r \rangle
\nonumber \\
 && \psi^*_{\bm{p'},m'} (\bm{k}_1, \sigma_1, ... \bm{k}_r ',\sigma_r',...
 \bm{k}_n, \sigma_n)
  \psi_{\bm{p}, m} (\bm{k}_1,\sigma_1, ... \bm{k}_r,\sigma_r, ... \bm{k}_n,
  \sigma_n) \nonumber \\
 &&  \delta^{(3)}(\bm{p}' - \bm{k}_1 - \bm{k}_2...-\bm{k}'_r ...-
\bm{k}_n) \delta^{(3)}(\bm{p} - \bm{k}_1 - \bm{k}_2 ...-\bm{k}_r
... - \bm{k}_n) .
 \label{eq.5.13}
\end{eqnarray}
After some manipulation this can be written as:
\begin{eqnarray}
\langle \bm{p}', m' | J_i | \bm{p}, m \rangle & = & (2\pi)^3
\,2p_{0} \sum_n \sum_{\sigma, \sigma'} \int d^3k \, d^3 k' \,
\delta^{(3)}(\bm{p'} - \bm{p} + \bm{k} - \bm{k'}) \,
\rho^{m' \, m}_{\sigma'\, \sigma}(\bm{k'},\bm{k}) ^{a} \nonumber \\
& & \frac{1}{\sqrt{(2\pi)^3 2k'_0}} \langle \bm{k'}, \sigma' |J_i
|\bm{k},  \sigma \rangle \frac{1}{\sqrt{(2\pi)^3 2k_0}}
\label{eq.5.14}
\end{eqnarray}
where we have introduced a density matrix for the internal motion
of type `$a$' partons in a proton of momentum $ \gb{p}$ :
\begin{eqnarray}
\rho^{m' \, m}_{\sigma' \, \sigma}(\bm{k'},\bm{k})^a & \equiv &
\sum_{n, r(a)} \sum_{\sigma_i} \sum_{ \sigma'_r} \delta_{\sigma \,
\sigma_r} \delta_{\sigma'\, \sigma'_r} \nonumber \\
& & \int d^3k'_r \, d^3k_1...d^3k_r...d^3k_n \, \delta
^{(3)}(\bm{k} -
\bm{k_r}) \, \delta^{(3)}(\bm{k'} - \bm{k'_r})\nonumber \\
& & \psi ^*_{p'm'}(\bm{k_1}, \sigma_1,...\bm{k'_r},
\sigma'_r,...\bm{k_n},\sigma_n ) \,
\psi_{pm}(\bm{k_1},\sigma_1,...\bm{k_r}, \sigma_r,...\bm{k_n},
\sigma_n ) \nonumber \\
& & \delta^{(3)}(\bm{p} - \bm{k_1}\  ...-\bm{k_r} \ ...-\bm{k_n})
\label{eq.5.15}
\end{eqnarray}
 Here $a$, which we will frequently suppress,
denotes the type of parton: quark, anti-quark or gluon. The sum
goes over all Fock states and, within these states, over the spin
and momentum labels $r$
 corresponding to the parton type $a$. Eqs.~(\ref{eq.5.14}) and (\ref{eq.5.15}) are the basis for
 the angular momentum sum rules.

The two terms in Eq.~(\ref{J11}) applied to the parton matrix
elements in Eq.~(\ref{eq.5.14}) suggest a spin part and an orbital
 part for quarks and gluons. This decomposition is a little misleading at this level, as we will see,
 but we will use it here to organize the various pieces. First consider the spin part of the matrix
 element when $k$ is the momentum carried by a quark.
\begin{eqnarray}
\langle \bm{p}', m' | J_i | \bm{p},m
 \rangle^{\rm quark spin} & = &
 (2\pi)^3 \,2p_0\, \delta^{(3)}(\bm{p}' - \bm{p})
  \int d^3 k d^3 k'\; \delta^{(3)}(\bm{k}-\bm{k}')
\nonumber \\
 &&
  \times \sum_{\sigma,\sigma'} \half(\bm{\sigma}_i)_{\sigma '\, \sigma } \;
 \rho_{\sigma'\,\sigma}^{m' \, m}(\bm{k'},\bm{k})^q ,
  \label{eq.5.18}
\end{eqnarray}
where here $\bm{\sigma}_i$ denotes the Pauli spin matrix of
Eq.~(\ref{J11}).

The spin part for the gluons is completely analogous, but now
$\sigma$ and $\sigma'$ in Eq.~(\ref{eq.5.14}) refer to the gluon
helicity $\lambda$. From Eq.~(\ref{eq.5.041}), which is diagonal
in helicity, we obtain
\begin{eqnarray}
\langle \bm{p}', m' | J_i | \bm{p},m
 \rangle^{\rm gluon spin} & = &
 (2\pi)^3 \,2p_0\, \delta^{(3)}(\bm{p}' - \bm{p})
  \int d^3 k d^3 k'\; \delta^{(3)}(\bm{k}-\bm{k}')
\nonumber \\
 &&
\eta_i \, \lambda\; \rho_{\lambda \, \lambda}^{m'\,
m}(\bm{k'},\bm{k})^G \label{eq.5.19}
\end{eqnarray}

The orbital part is somewhat different because of the derivative
of the $\delta$-function that enters, and we have stressed the
need for a proper wave packet treatment, as carried out in
Section~5. However in Eq.~(\ref{eq.5.13}) or (\ref{eq.5.14}) the
partons are not in plane wave states and the partonic wave
function $\psi$ plays the role of a wave packet. Thus we may
proceed directly by inserting the orbital piece of Eq.~(\ref{J11})
into Eq.~(\ref{eq.5.13}).

 We have
\begin{eqnarray}
 \langle \bm{p}', m' | J_i | \bm{p}, m
 \rangle^{\rm orbital} & = &
 (2\pi)^3 2 p_0 \sum_n  \sum_{\{\sigma\}} \sum_r
  \int d^3k'_r \, d^3 k_1 \dots d^3 k_r  \dots d^3 k_n
 \sqrt{\frac{k^0_r}{k^{\prime 0}_r}}
\nonumber \\
 &&
 \psi^*_{\bm{p'}, m'}(\bm{k}_1, \sigma_1, ..., \bm{k}'_r,\sigma_r, ...
 \bm{k}_n, \sigma_n)\, \psi_{\bm{p}, m}(\bm{k}_1, \sigma_1, ...,\bm{k}_r,\sigma_r, ... \bm{k}_n,\sigma_n )
\nonumber \\
 &&
 \delta^{(3)}(\bm{p}' - \bm{k}_1 - ... - \bm{k}'_r - ...  \bm{k}_n)
 \delta^{(3)}(\bm{p} - \bm{k}_1 - ... - \bm{k}_r - ...  \bm{k}_n)
\nonumber \\
 & &
i (\bm{k}_r\times\bm{\nabla}_{k_r})_i \delta^{(3)}(\bm{k}'_r -
\bm{k}_r) \label{eq.A}
\end{eqnarray}
Integrating over $\bm{k}_r$ by parts yields
\begin{eqnarray}
 \langle \bm{p}', m' | J_i | \bm{p}, m
 \rangle^{\rm orbital} & = &
 - (2\pi)^3 2 p_0 \sum_n \sum_{\{\sigma\}} \sum_r
  \int d^3k'_r \, d^3 k_1\dots d^3 k_r \dots d^3 k_n
 \sqrt{\frac{k^0_r}{k^{\prime 0}_r}}
\nonumber \\
 &&
 \psi^*_{\bm{p'}, m'}(\bm{k}_1, \sigma_1, ..., \bm{k}'_r,\sigma_r, ...
 \bm{k}_n, \sigma_n)
\nonumber \\
 &&
 \delta^{(3)}(\bm{p}' - \bm{k}_1 ... - \bm{k}'_r ...  -\bm{k}_n)
 \delta^{(3)}(\bm{k}'_r - \bm{k}_r)
 \\
 & & i (\bm{k}_r \times\bm{\nabla}_{k_r})_i
 [\psi_{\bm{p}, m}(\bm{k}_1, \sigma_r.., \bm{k}_r,\sigma_r..\bm{k}_n,\sigma_n )
 \delta^{(3)}(\bm{p} - \bm{k}_1 ..- \bm{k}_r .. -\bm{k}_n)]
 \nonumber
\label{eq.B}
\end{eqnarray}

The derivative produces two terms. The one arising from the
derivative of the delta-function is

\begin{eqnarray}
 & = & -(2\pi)^3 2 p_0 \sum_n \sum_{\{\sigma\}}
 \int d^3 k_1 \dots \dots d^3 k_n
 \psi^*_{\bm{p}, m'}(\bm{k}_1, \sigma_1, ..., \bm{k}_n,\sigma_n
 )\, \psi_{\bm{p}, m}(\bm{k}_1, \sigma_1, ...,
 \bm{k}_n,\sigma_n)
\nonumber \\
& & \delta^{(3)}(\bm{p}' - \bm{k}_1 ... - \bm{k}_n)
 \sum_r i (\bm{k}_r\times\bm{\nabla}_{k_r})_i
 \delta^{(3)}(\bm{p} - \bm{k}_1 ... - \bm{k}_n)
 \label{eq.C}
 \end{eqnarray}

Now it is easy to check that
\begin{equation}
\sum_r i (\bm{k}_r\times\bm{\nabla}_{k_r})_i
 \delta^{(3)}(\bm{p} - \bm{k}_1 - ...-\bm{k}_r-... - \bm{k}_n) =
 - (\bm{p}\times\bm{\nabla}_p)_i \delta^{(3)}(\bm{p} - \bm{k}_1 - ... - \bm{k}_n)
\label{eq.D}
\end{equation}
and putting this into Eq.~(\ref{eq.C}) and using the normalization
and orthogonality of the wave functions, this term produces

\begin{equation}\label{eq.F}
2p_0 (2\pi)^3 i\epsilon_{ijk}p_j\frac{\partial}{\partial
p_k}\delta^{(3)} (\bm{p'} - \bm{p} )\, \delta_{m m'}
\end{equation}
which just cancels the derivative of the delta-function in
Eq.~(\ref{J11}).

The other term in the differentiation in Eq.~(\ref{eq.B}) yields
\begin{equation}\label{eq.E}
2p_0 (2\pi)^3 \delta^{(3)} (\bm{p'} -\bm{p} )\langle L_i \rangle^a
_{m'\, m}
\end{equation}
where $\langle L_i \rangle^a_{m' \, m} $ is the contribution from
the internal angular momentum arising from partons of type $a$,
given by

\begin{eqnarray}
\langle L_i \rangle^a_{m' \, m}& = & \sum_n \sum_{ \{\sigma \}}
 \int d^3 k_1 \dots \dots d^3 k_n
\psi^*_{\bm{p}, m'}(\bm{k}_1, \sigma_1, ..., \bm{k}_n, \sigma_n)
\nonumber \\
  & & \sum_{r(a)} \{ [-i (\bm{k}_r\times\bm{\nabla}_{k_r})_i]
 \psi_{\bm{p}, m}(\bm{k}_1, \sigma_1, ...\bm{k}_r,\sigma_r, ... \bm{k}_n,\sigma_n
 ) \} \nonumber \\
& & \delta^{(3)}(\bm{p} - \bm{k}_1 - ... - \bm{k}_n) \label{eq.F1}
\end{eqnarray}
where the sum over $r(a)$ means a sum over those $r$-values
corresponding to partons of type $a$. Note that $a$ can refer to
both quarks and gluons; the  structure of Eq.~(\ref{eq.F1}) is the
same for both. Note also that the orbital angular momentum defined
in this way is not in general the same as that given by the matrix
element $M_{\rm orb}^{ij}$, Eqs.~(\ref{eq.3.54}) and
(\ref{eq.3.71}), which contain gluon spin-dependent parts which
here are included in the spin part of the matrix elements of
$J_i$. This difference is important for transverse polarization
but not for longitudinal polarization. Furthermore, it is
important to realize that the
 orbital angular momentum defined in this way depends on the basis states used for the partons. In particular,
  because of the momentum dependence of the transformation from canonical to helicity
  basis, Eq.~(\ref{helicitytocanonical}).
\beq \rho_{\lambda'\,\lambda}^{{m' \,m}}(\bm{k'},\bm{k}) =
\mathcal{D}^{1/2}_{m'_{r} \lambda' }(R(\bm{k}'))^*
 \rho_{m'_r\,
m_r}^{m' \,m}(\bm{k'},\bm{k})\ \mathcal{D}^{1/2}_{m_{r} \lambda
}(R(\bm{k})) \eeq
the orbital angular momentum will not be the same in the two
bases; cf. Eqs.~(\ref{helicity3}) and (\ref{helicity4}). Of
course, for $k_{z} \rightarrow \infty$ with finite $k_{x}, k_{y}$
the $z$-component will be the same but the other components will
not be.

In this discussion we have used a fixed-axis quantization for the
quarks
 since it has a more transparent connection to the polarization states of
  the proton, which in most cases are more naturally described in that way.
  At the same time, because the gluons are massless it is most natural to
   use helicity states to describe them, so we are led to a mixed notation.
   There is no real problem with this, but there may be occasions where one
    wishes to treat the quarks in helicity states as well.

Putting Eqs.~(\ref{eq.E}), (\ref{eq.5.18}) and (\ref{eq.5.19})
into Eq.~(\ref{eq.5.14}) , utilizing Eq.~(\ref{J11}) for its LHS ,
and cancelling the factors $2p_0 (2\pi)^3 \delta( \bm{p'} -
\bm{p})$, we end up with the general sum rule for a spin $\half$
nucleon:
\begin{eqnarray}\label{eq.G}
\half (\bm{\sigma}_i)_{m'\, m}& =& \int d^3 \bm{k}\ \left[
\half(\bm{\sigma}_i)_{\sigma' \, \sigma }\, \ \rho^{m'
\,m}_{\sigma' \, \sigma} (\bm{k}, \bm{k} )^{q + \bar{q}}+
\lambda \ \eta_i(\bm{k}) \ \rho^{m' \, m}_{\lambda \,
\lambda} (\bm{k}, \bm{k})^G \right] \nonumber \\
& & + \langle L_i\rangle^{q+ \bar{q}}_{m' \, m} + \langle
L_i\rangle^G_{m' \, m}
\end{eqnarray}
where $\eta_i$ is given in Eq.~(\ref{helicity2}).

\subsection{A new sum rule}

For proton matrix elements of $J_z$,  Eq.~(\ref{eq.G}) is
non-vanishing only for $m'=m$. It then becomes the classic sum
rule Eq.~(\ref{eq.5.07}). There is one other independent sum rule
that can be  obtained from this general one; one way to obtain it
is to consider the matrix elements of $J_x$ which are
non-vanishing only for $m'=-m$. The LHS of Eq.~(\ref{eq.G}) is
then equal to $\half$. The quark spin contribution to the RHS (an
identical expression holds for the antiquarks) is
\begin{equation}\label{eq.I}
\half \int d^3 \bm{k} \half \left[  \rho^{+ \, -}_{+ \, -} +
\rho^{+ \, -}_{- \, +} + \rho^{- \, +}_{- \, +} + \rho^{- \, +}_{+
\, -} \right]^q
\end{equation}
where +/- refers to $\pm \half$.
By rotating the system through $\pi$ about the $z$-axis, it is
easy to see that elements of $\rho^{m',m}_{\sigma',\sigma}$ with
$(-1)^{m-m' -\sigma +\sigma'}=-1$ are odd under this rotation and
so will integrate to zero when integrated over $\bm{k_T}$. This
enables us to rewrite the expression (\ref{eq.I}),  the quark
contribution, in a way that has a nice interpretation, viz.
\begin{equation}\label{eq.I'}
\half \int d^3 \bm{k} \, \half \left[ \rho^{+ \, +}_{+ \,-} +
\rho^{+ \,+}_{-\, +} + \rho^{-\, -}_{+\, -} + \rho^{- \, -}_{- \,
+} + \rho^{+ \, -}_{+ \, -} + \rho^{+ \, -}_{- \, +} + \rho^{- \,
+}_{- \, +} + \rho^{- \, +}_{+ \, -} \right]^q.
\end{equation}

Consider the proton state with spin oriented along OX,
perpendicular to the direction of motion
\begin{equation}\label{eq.H}
| \bm{p}, \bm{s}_x \rangle = \frac{1}{\sqrt{2}}\{ | \bm{p}, m=1/2
\rangle + |\bm{p}, m=-1/2 \rangle \}
\end{equation}
To understand the content of expression (\ref{eq.I'}) write
schematically
\begin{equation}\label{eq.J}
\rho^{m' \, m}_{\sigma' \, \sigma} = \sum_{X=all} \psi^*_{m'}
(\sigma', X )\, \psi_{m} (\sigma, X )
\end{equation}
Now the number density of quarks with spin along or opposite to
$OX$, denoted by $\pm \hat{\bm{s}}_x $ in a proton spinning along
$OX$ is
\begin{equation}\label{eq.K}
q_{\pm \hat{\bm{s}}_x /\bm{s}_x} (\bm{k}) = \sum_{X=all}
|\psi_{\bm{s}_x }( \pm \hat{\bm{s}}_x, X ) |^{\,2}
\end{equation}
where
\begin{equation}\label{eq.L}
\psi_{\bm{s}_x} ( \pm \hat{\bm{s}}_x )= \half [ \psi_{+} (+) \pm
\psi_{+} (-) + \psi_{-}(+) \pm\psi_{-} (-) ]
\end{equation}
so that (suppressing the $\sum_{X=all}$)
\begin{equation}\label{eq.M}
q_{\hat{\bm{s}_x}/\bm{s}_x} (\bm{k}) -
q_{-\hat{\bm{s}_x}/\bm{s}_x}(\bm{k}) = Re \{\, [ \,\psi_{+}(+) +
\psi_{-} (+)\, ]^* [\, \psi_{-} (-) + \psi_{+} (-)\, ]\, \}
\end{equation}
which, via Eq.~(\ref{eq.J}), is exactly the integrand in
Eq.~(\ref{eq.I'}). Thus the expression (\ref{eq.I}) is equal to
\begin{equation}\label{eq.N}
\half \int d^3 \bm{k}\, [ \, q_{\hat{\bm{s}_x}/\bm{s}_x} (\bm{k})
- q_{-\hat{s}_x/\bm{s}_x} (\bm{k})\, ]=\half \int dx \, d^2
\bm{k}_T\, [ \, q_{\hat{\bm{s}_x}/\bm{s}_x} (x, \bm{k}_T) -
q_{-\hat{s}_x/\bm{s}_x} (x, \bm{k}_T)\, ]
\end{equation}
and there is an an analogous term for the antiquarks.

Now it is known \cite{ref.13} that
\begin{eqnarray}
 q^a_{\hat{s}_x/s_x} (x, \bm{k}_T) -
 q^a_{-\hat{s}_x/s_x} (x, \bm{k}_T)  & =  &
\Delta'_T q^a(x, k^2_T)
 \nonumber \\
 & &  +
 \cos2\phi \frac{k^2_T}{2M^2}
 h^{\perp a}_{1T} (x, k^2_T)
 +\sin\phi \frac{k_T}{M} h^{\perp a}_1 (x, k^2_T)
 \nonumber \\
\label{eq.5.59}
\end{eqnarray}
where $\phi$ is the azimuthal angle of $\bm{k}_T$ , $\Delta'_T
q^a(x, k^2_T)$ is the same as $h^a_1 (x, k^2_T)$ in the notation
of Ref.~\cite{ref.11}, and
\begin{equation}
 \Delta_T q^a (x) \equiv h^a_1 (x) =
 \int d^2 \bm{k}_T \Delta'_T q^a(x, k^2_T).
\label{eq.5.60}
\end{equation}

Substituting Eq.~(\ref{eq.5.59}) into Eq.~(\ref{eq.N}) and
integrating over the direction of $\bm{k}_T$ , we end up with the
quark contribution to the RHS of Eq.~(\ref{eq.G}):
\begin{equation}
 \half \int dx \,h_1(x) = \half \int dx \sum_{a,\bar{a}} h^a_1(x)
 \equiv \half \int dx \sum_{a,\bar{a}} \Delta_T q^a (x).
\label{eq.5.61}
\end{equation}

We turn now to the gluon contribution to the RHS of
Eq.~(\ref{eq.G}), which is
\begin{equation}\label{eq.O}
\int d^3 \bm{k} \, \eta_x(\bm{k}) \, (\half \, [\rho^{+ \,+ }_{1
\, 1 } -\rho^{+ \, + }_{-1 \, -1 } + \rho^{- \, - }_{1 \, 1 } -
\rho^{ - \, - }_{-1 \, -1 } + \rho^{+ \, - }_{1 \, 1 } - \rho^{+
\,- }_{-1 \, -1 } + \rho^{- \, + }_{1 \, 1 } - \rho^{- \, + }_{-1
\, -1 } \, ] ),
\end{equation}
where $\pm 1$ refers to the gluon helicity. Once again we have
added in terms which integrate to zero in order to get a nice
interpretation in terms of densities. (Recall that $\eta_x$
contains the factor $\cos\phi$, and the factor
$\rho^{m',m}_{\lambda'\,\lambda}$ with $(-1)^{m' - m -\lambda'
+\lambda}=+1$ are even under $\phi \rightarrow \pi \pm \phi$.)

Now consider
\begin{eqnarray}\label{eq.P}
\Delta G_{h/\bm{s}_x} & \equiv &  G_{1/\bm{s}_x} - G_{-1/\bm{s}_x}
\nonumber \\
& =&  \sum_{X=all} \{ |\psi_{\bm{s}_x} ( 1 ,X )|^2 -
|\psi_{\bm{s}_x} (-1, X)|^2 \}
\end{eqnarray}
Carrying out the analogue of Eq.~(\ref{eq.L}) we find that the RHS
of Eq.~(\ref{eq.P}) is exactly equal to the terms in parenthesis
in Eq.~(\ref{eq.O}). Thus the gluon contribution to the RHS of
Eq.~(\ref{eq.G}) is
\begin{equation}\label{eq.Q}
\int d^3 \bm{k}\, \eta_x(\bm{k}) \, \Delta G_{h/\bm{s}_x}(\bm{k})
= \int dx \,d^2 \bm{k}_T \, \eta_x (\bm{k}) \, \Delta
G_{h/\bm{s}_x} (x, \, \bm{k}_T)
\end{equation}

It is easy to see, geometrically, that $\Delta G_{ h/\bm{s}_x}(x,
\bm{k}_{T}) $ contains a factor $k_x$ and we make this
 explicit by writing \cite{ref.11}
 \beq
 \Delta G_{ h/\bm{s}_x}(x, \bm{k}_{T}) = \frac{k_x}{M}
  g_{1 T}^G (x, k^2_T) \label{g1T}.
\eeq Then the contribution of the gluon spin to the RHS of
Eq.~(\ref{eq.G}) i.e. to the proton whose spin is in the
$x$-direction is
\begin{eqnarray}
\Delta G_{h/\bm{s}_x} & = & \int dx \,d^2 \bm{k}_{T}
\eta_x \frac{k_x}{M} g_{1 T}^G (x, k^2_{\perp}) \nonumber  \\
&=& \pi \int dx \, k_T dk_T \frac{\sqrt{x^2 p^2 + k_{T}^2}-x\
p}{M} g^G_{1T}(x, k_{T}^2)). \label{eq.R}
\end{eqnarray}
where we have used Eq.~(\ref{helicity2}). As $p \rightarrow
\infty$ this piece vanishes and so the gluon spin does not
contribute to the transverse spin sum rule.

Finally, the internal orbital angular momentum terms $\langle L_x
\rangle^{q}_{\bm{s}_x} $ and $ \langle L_x \rangle^{G}_{\bm{s}_x}$
are obtained from Eq.~(\ref{eq.F1}) by the replacement
\begin{equation}\label{eq.S}
\psi_{\bm{p},\, m} \rightarrow \psi_{\bm{p}, \, \bm{s}_x} =
\frac{1}{\sqrt{2}}\, \left[ \psi_{\bm{p}, \, \half} +
\psi_{\bm{p},\, -\half} \right]
\end{equation}

Putting together the various pieces of the RHS of Eq.~(\ref{eq.G})
we obtain a new, transverse spin sum rule. Since the same result
holds when considering $J_y$ with the proton polarized along $OY$,
we prefer to state the result in the more general form: for a
proton in an eigenstate of transverse spin with eigenvector along
$\bm{s}_T$
\begin{equation}\label{eq.T}
\half = \half\, \sum_{q, \,\bar q }\, \int dx \, \Delta _T q^a (x)
+ \sum_{q, \, \bar q, \, G }\langle L_{\bm{s}_T} \rangle^a
\end{equation}
where $L_{\bm{s}_T}$ is the component of $\bm{L}$ along
$\bm{s}_T$. No such sum rule is possible with the Jaffe-Manohar
formula because, as $p \rightarrow \infty$, Eq.~(\ref{eq.5.05})
for $i=x,y$ diverges.

This has a very intuitive appearance, very similar to
Eq.~(\ref{eq.5.07}), but there are a couple of points to be made.
First of all, this sum rule is distinct from the tensor charge
$\delta$ sum rule of Jaffe and Ji  \cite{Jaffe & Ji} because here
the quarks and anti-quarks {\em add} whereas  in $\delta$ the
difference enters. It is possible that this difference may be of
some use in disentangling the quark and anti-quark transverse spin
structure functions. Second, the integral does not correspond to
the matrix element of  the axial vector current, which via
Eq.~(\ref{eq.3.70}), vanishes for $p \rightarrow \infty$ when
$\bm{p} \cdot \bm{s}=0$.

The structure functions $\Delta_T q^a (x) \equiv h^q_1(x)$ are
most directly measured in doubly polarized Drell-Yan reactions
where the asymmetry is proportional to
\begin{equation}
 \sum_a e^2_a [\Delta_T q^a (x_1) \Delta_T \bar{q}^a (x_2) + (1 \leftrightarrow 2)].
\label{eq.5.64}
\end{equation}
For a detailed discussion see Ref.~\cite{ref.13}.

They can also be determined from the asymmetry in semi-inclusive
hadronic interactions like
\[
 p + p(\bm{s}_T) \to H + X
\]
where $H$ is a detected hadron, typically a pion, \cite{ref.17},
and in SIDIS reactions with a transversely polarized target
\cite{ref.18, ref.19}
\[
 \ell + p(\bm{s}_T) \to \ell + H + X.
\]
The problem is that in these semi-inclusive reactions $\Delta_T
q^a (x)$ always occurs multiplied by a function which depends on
the largely unknown Collins fragmentation function. But progress
is being made and we expect to have a reasonable estimate of
$\Delta_T q^a (x)$ soon \cite{ref.20}.

\section{Summary}
The standard derivation of the tensorial structure of the
expectation value of the angular momentum $\bm{J}$, for a
relativistic spin-$s$ particle, in which the matrix elements of
the angular momentum operators are related to the matrix elements
of the energy-momentum tensor, is rendered difficult by the
singular nature of the operators involved. The fact that the
operators are space integrals of densities that contain explicit
factors of $x^{\mu}$ means that the densities do not transform
like local operators (we have called these {\em compound}
operators) and has the consequence that the matrix elements of
$\bm{J}$ are highly singular, containing derivatives of
delta-functions.  Consequently the evaluation of the expectation
values of $\bm{J}$ requires a careful limiting procedure,
beginning with the off-diagonal elements $\la
p',\sigma|\bm{J}|p,\sigma \ra$, where $\sigma$ labels the kind of
spin state under consideration. We have shown that the results in
the literature are incorrect, and have derived the correct
expressions in three different ways, two of them based on a
careful wave-packet treatment of the standard approach, and the
third, quite independent, based on the known rotational properties
of the spin states, which circumvents the use of the
energy-momentum tensor. All three methods yield the same results,
given in Eqs.~(\ref{eq.5.011}), (\ref{J11}) and (\ref{eq.5.041}).
We have emphasized that these matrix elements are not Lorentz
tensors, but nevertheless the resulting forms have the correct
Lorentz transformation properties when taken in conjunction with
the corresponding forms for the matrix elements of the boost
operators. We have given some attention to the relation between
the matrix elements for canonical spin states and  helicity
states, which is not obvious, and requires some care because of
the derivatives of the delta functions which enter.
 Using a Fock-space picture of the proton, we have used our results to obtain a new sum
rule for a transversely polarized nucleon, Eq.~(\ref{eq.T1}),
which involves the transverse spin or  transversity distribution
$\Delta_Tq(x)\equiv h_{1}(x)$, and which is similar in form to the
classic longitudinal spin sum rule.

\section{Acknowledgements}
    E.L. is grateful to Professor Piet Mulders and to Dr Sally Dawson
    for their hospitality at the Vrije University, Amsterdam and at
    Brookhaven National Laboratory, respectively. This research project
    was  supported by the Royal Society of Edinburgh Auber
    Bequest and by
    the Foundation for Fundamental Research on Matter (FOM) and the
    Dutch Organization for Scientific Research (NWO), and was authored under
    Contract No. DE-AC02-98CH10886 with the U.S. Department of Energy.
    Accordingly, the U.S.
    Government retains a non-exclusive, royalty-free license to publish or
    reproduce the published form of this contributions, or allow others to do so,
    for U.S. Government purposes.

\end{document}